\documentclass[10pt]{article} % For LaTeX2e
\usepackage[accepted]{tmlr}
\usepackage{url}

\usepackage[breaklinks]{hyperref}
\usepackage{url}
\usepackage{CJKutf8}
\usepackage{graphicx}
\usepackage{longtable}
\usepackage{array}
\usepackage{booktabs}
\usepackage{multirow}
\usepackage[normalem]{ulem}
\usepackage{comment}
\usepackage[most]{tcolorbox}
\definecolor{whitesmoke}{HTML}{F5F5F5}
\colorlet{background}{whitesmoke}
\usepackage{enumitem}
\usepackage{float}
\usepackage{titlesec}
\titlespacing{\subsubsection}{4pt}{*0}{*0}

\newtcolorbox{researchbox}{%
breakable,boxrule=1pt,enhanced jigsaw, sharp corners,pad at break*=1mm,colbacktitle=background,colback=background,colframe=black,coltitle=black,toptitle=2mm,bottomtitle=1.5mm,width=\linewidth,fonttitle=\bfseries,parbox=false,title=Example Research
Questions,boxed title style={empty,boxrule=0pt, bottom=0pt}
}

\title{Open Problems in Technical AI Governance}

\author{\name Anka Reuel\thanks{Equal contribution; corresponding authors; order randomized. Work completed while BB was at Centre for the Governance of AI.\\Given its scope, inclusion as an author does not entail endorsement of all aspects of the paper, with the exception of AR and BB.\\Cite as Reuel, Bucknall, et al. (2025) ``Open Problems in Technical AI Governance.''} $^1$ \email anka.reuel@stanford.edu \\ 
      \name Ben Bucknall\footnotemark[1] $^{2,3}$ \email bucknall@robots.ox.ac.uk
      \AND
      \name Stephen Casper$^4$,
      \name Tim Fist$^{5,6}$,
      \name Lisa Soder$^{7}$,
      \name Onni Aarne$^{8}$,
      \name Lewis Hammond$^{2,9}$,
      \name Lujain Ibrahim$^{2}$,\\
      \name Alan Chan$^{10,11}$,
      \name Peter Wills$^{2,10}$,
      \name Markus Anderljung$^{10}$,
      \name Ben Garfinkel$^{10}$,
      \name Lennart Heim$^{10}$,\\
      \name Andrew Trask$^{2,12}$,
      \name Gabriel Mukobi$^{1}$,
      \name Rylan Schaeffer$^{1}$,
      \name Mauricio Baker$^{13}$,
      \name Sara Hooker$^{14}$,\\
      \name Irene Solaiman$^{15}$,
      \name Alexandra Sasha Luccioni$^{15}$,
      \name Nitarshan Rajkumar$^{16}$,
      \name Nicolas Mo\"es$^{17}$,\\
      \name Jeffrey Ladish$^{18}$,
      \name David Bau$^{19}$,
      \name Paul-Andrei Bricman$^{20}$,
      \name Neel Guha$^{1}$,
      \name Jessica Newman$^{21}$,\\
      \name Yoshua Bengio$^{11,22}$,
      \name Tobin South$^{23}$,
      \name Alex Pentland$^{24}$,
      \name Sanmi Koyejo$^{1,25}$,
      \name Mykel J. Kochenderfer$^{1}$,\\
      \name Robert Trager$^{2,3}$
      \AND
      \addr $^1$Stanford University $^2$University of Oxford $^3$Oxford Martin AI Governance Initiative $^4$MIT CSAIL\\
      $^5$Institute for Progress $^6$Center for a New American Security\\
      $^7$interface - Tech Analysis and Policy Ideas for Europe e.V. $^8$Institute for AI Policy and Strategy\\
      $^9$Cooperative AI Foundation $^{10}$Centre for the Governance of AI $^{11}$Mila $^{12}$OpenMined $^{13}$Independent Researcher\\
      $^{14}$Cohere for AI $^{15}$Hugging Face $^{16}$University of Cambridge $^{17}$The Future Society $^{18}$Palisade Research\\
      $^{19}$Northeastern University $^{20}$Noema Research $^{21}$University of California, Berkeley $^{22}$University of Montreal $^{23}$MIT\\
      $^{24}$Stanford HAI $^{25}$Virtue AI
      }

\def\month{04}  % Insert correct month for camera-ready version
\def\year{2025} % Insert correct year for camera-ready version
\def\openreview{\url{https://openreview.net/forum?id=1nO4qFMiS0}} % Insert correct link to OpenReview for camera-ready version

\begin{document}

\maketitle

\begin{abstract}
AI progress is creating a growing range of risks and opportunities, but it is often unclear how they should be navigated. In many cases, the barriers and uncertainties faced are at least partly technical. Technical AI governance, referring to technical analysis and tools for supporting the effective governance of AI, seeks to address such challenges. It can help to (a) identify areas where intervention is needed, (b) assess the efficacy of potential governance actions, and (c) enhance governance options by designing mechanisms for enforcement, incentivization, or compliance. In this paper, we explain what technical AI governance is, outline why it is important, and present a taxonomy and incomplete catalog of its open problems. This paper is intended as a resource for technical researchers or research funders looking to contribute to AI governance.  
\end{abstract}

\makeatletter
\let\origsection\section
\renewcommand\section{\@ifstar{\starsection}{\nostarsection}}

\newcommand\nostarsection[1]
{\sectionprelude\origsection{#1}\sectionpostlude}

\newcommand\starsection[1]
{\sectionprelude\origsection*{#1}\sectionpostlude}

\newcommand\sectionprelude{%
  \vspace{1em}
}

\newcommand\sectionpostlude{%
  \vspace{1em}
}
\makeatother

\newpage
\section{Introduction}\label{1-introduction}
The rapid development and adoption of artificial intelligence (AI)
systems\footnote{Our understanding of AI systems follows that of
  \citep{Basdevant2024-qg}, encompassing infrastructure
  such as compilers, model components such as datasets, code, and
  weights, as well as UX considerations.} has prompted a great deal of
governance action from the public sector,\footnote{See, for example, \begin{CJK*}{UTF8}{gbsn}\citep{The_White_House2023-ru,The_White_House_Office_of_Science_and_Technology_Policy2023-ia,Presidency_of_the_Council_of_the_European_Union2024-ty,Department_for_Science_Innovation_and_Technology2023-hh,Department_for_Science_Innovation_and_Technology2023-qz,Advisory_Body_on_Artificial_Intelligence2023-ma,European_Commission2023-db,2023-vo}\end{CJK*}}
academia and civil society
\citep{Anderljung2023-pa,Moes2023-qa,Barrett2023-kg}, and
industry
\citep{Anthropic2023-cv,Microsoft2023-ov,Dragan2024-rm,OpenAI2024-cu},
with the aim of addressing potential risks while capitalizing on
benefits.

However, key decision-makers seeking to govern AI often have insufficient information for identifying the need for intervention and assessing the efficacy of different governance options. Furthermore, the technical tools necessary for successfully implementing governance proposals are often lacking \citep{Reuel2024-ag}, leaving uncertainty regarding how policies are to be implemented. For example, while the concept of watermarking\footnote{Watermarks are signals placed
  in output content that are imperceptible to humans, but easily
  detectable through application of a specific \emph{detector} algorithm.}
AI-generated content has gained traction among policymakers
\citep[see for example][]{Council_of_the_European_Union2024-pt,The_White_House2023-yx,G7_leaders2023-lu,Department_for_Science_Innovation_Technology2023-xm},
it is unclear whether current methods are sufficient for achieving
policymakers' desired outcomes, nor how future-proof such methods will
be to improvements in AI capabilities
\citep{Zhang2023-qg,Ghosal2023-nl}. Addressing these and similar issues will require further targeted technical advances.

In this paper, we aim to catalyse work aimed at addressing such challenges and uncertainties through motivating and introducing the field of \textbf{technical AI governance (TAIG)}, providing a taxonomy of subareas, and outlining open problems and research questions. Throughout the paper we use the term technical AI governance to refer to
% As such, in this paper we aim to provide an overview of \textbf{technical AI governance (TAIG)}, 
\emph{technical analysis and tools for supporting the effective
governance of AI}, where `technical' pertains to the physical sciences, mathematics, engineering, and related fields; and `AI governance' refers to the processes and structures through which decisions related to AI are made, implemented and enforced.\footnote{To the extent that this definition of TAIG
  includes measures for directly increasing the performance, safety, or
  robustness of AI systems, we only consider such measures for cases in
  which they support the governance of AI.}
AI governance includes research that is both non-technical (e.g. in politics, economics, law, philosophy, etc.) and technical in nature (e.g. in computer science, engineering, and mathematics). Through our use of this definition we simply refer to this latter category of work within the umbrella category of `AI governance.'

By the above definition, TAIG can contribute to AI governance in a number of
ways, such as by identifying opportunities for governance intervention,
informing key decisions, and enhancing options for implementation. For
example, deployment evaluations that assess the downstream impacts of a
system (see Section \ref{34-deployment}) could help identify a need for policy
interventions to address these impacts. Alternatively, being able to
design models that are robust to malicious modifications (see Section
\ref{64-deployment-3}) could add to the menu of governance options available to prevent
downstream misuse.

In particular, we make the following contributions:

\begin{itemize}
\item

  We introduce the field of technical AI governance and
  motivate the need for work in this area.

\item

  We present a taxonomy of TAIG arranged along two
  dimensions: \emph{capacities}, which refer to actions such as access
  and verification that are useful for governance, and \emph{targets},
  which refer to key elements in the AI value chain, such as data and
  models, to which capacities can be applied.

\item

  Finally, we outline open problems within each category of our
  taxonomy, along with concrete example questions for future research.

\end{itemize}

\begin{figure}
    \centering
    \includegraphics[width=\linewidth]{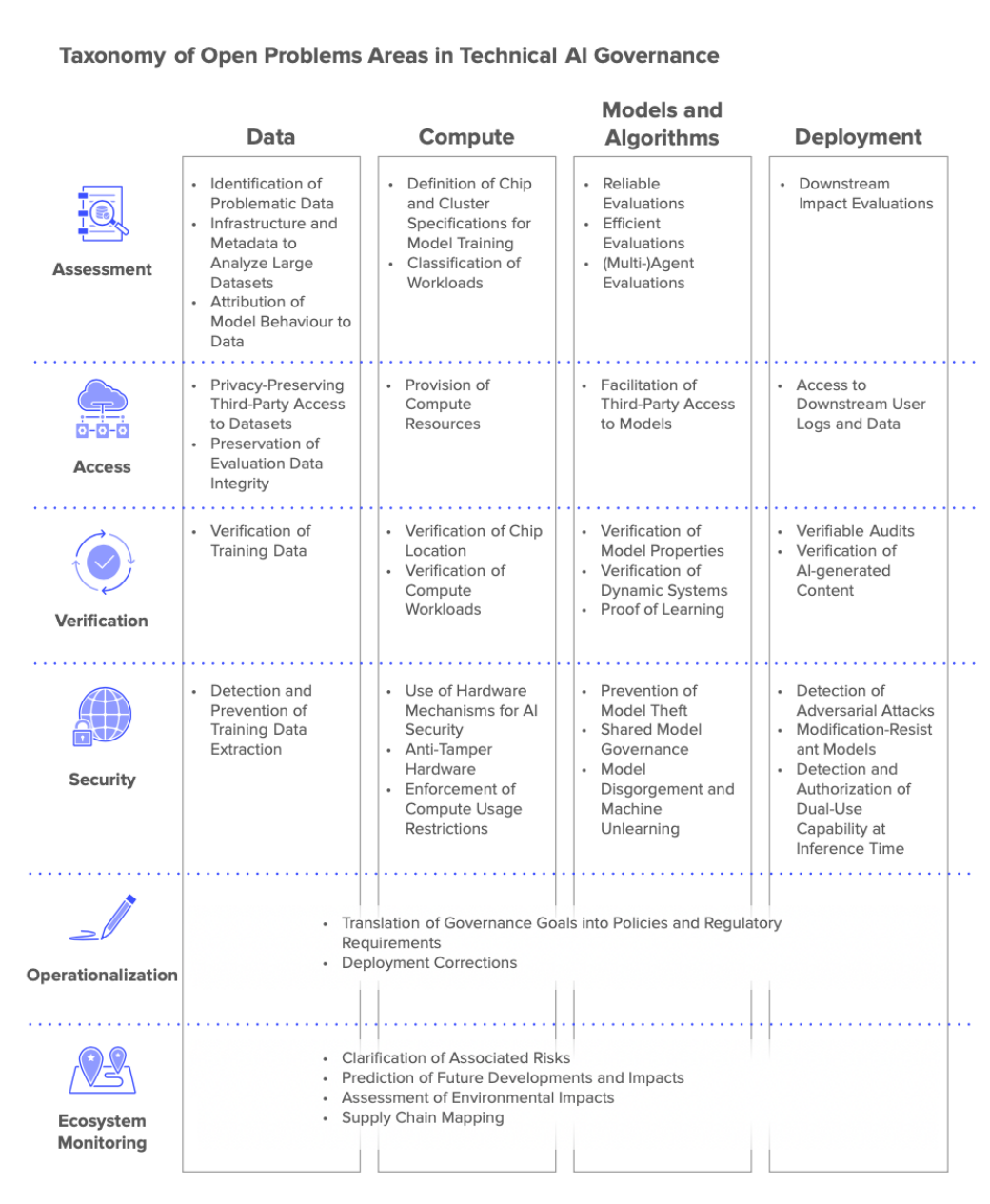}
    \caption{An overview of the open problem areas covered in this report, organized according to our taxonomy.}
    \label{fig:overview}
\end{figure}

Figure~\ref{fig:overview} provides an overview of the open problem areas, organized
according to the taxonomy. We hope that this paper serves as a resource
and inspiration for technical researchers aiming to direct their
expertise towards policy-relevant topics.

 % \newpage
 % \newpage

\subsection{Relation to AI
Governance}\label{11-relation-to-ai-governance}

As noted above, we define AI governance as the processes and structures through which
decisions related to AI are made, implemented, and enforced. It
encompasses the rules, norms, and institutions that shape the behavior
of actors in the AI ecosystem, as well as the means by which they are
held accountable for their actions.\footnote{For other proposed
  definitions of AI governance, see
  \citep{Bullock2022-ax,Daly2021-aa,Dafoe2018-fo}.} As
per our definition above, TAIG consists of \emph{technical analysis and
tools for supporting the effective governance of AI.} Here we outline
three ways in which TAIG can contribute to AI governance, which we refer to as
\emph{identifying}, \emph{informing}, and \emph{enhancing}.\footnote{A
  useful parallel to TAIG may be the concepts of regulatory technology
  (RegTech) and supervisory technology (SupTech) in the financial sector
  \citep{Bank_for_International_Settlements2021-va},
  which aim to support financial regulation and oversight.}

Firstly, \textbf{TAIG can identify areas where governance intervention
is needed}, through mapping technical aspects of AI systems to social and political concepts, typically conceived of as being addressed through governance.
For example, tracking and considering technical advances in AI video
generation could allow for more accurate predictions of the risk of
video deepfakes, and thus motivate the need for a governance response.

Secondly, \textbf{TAIG can inform governance decisions} by providing
decision-makers with more accurate information, allowing them to better
compare the effectiveness of different governance options. For example,
policymakers can choose between different regulatory instruments (for
example, registration or disclosure), as well as how they enforce
compliance (for example, \emph{ex ante} rules or \emph{post hoc}
adjudication), with the efficacy of these options potentially depending on technical details. Information could stem from implemented TAIG methods,
such as the outcome of assessments (see Section \ref{3-assessment}), or TAIG research
that maps or monitors the AI ecosystem (see Section \ref{8-ecosystem-monitoring}). For example,
more developed risk models for assessing potential harms of AI could inform
targeted policies for their mitigation.

Finally, \textbf{TAIG can enhance governance options} by providing or
enabling mechanisms for enforcing, incentivizing, or complying with
mandated requirements. For instance, developing methods for the robust
evaluation of models with black-box access could facilitate more
comprehensive third-party auditing, thereby enhancing enforcement of
safety requirements.

Previous overviews of AI governance, such as \citet{Dafoe2018-fo}'s AI governance research agenda, focus primarily on socio-political governance challenges, including labor displacement, inequality, shifts in national power, and AI race dynamics, or taxonomizes themes, gaps, and issues in AI governance more broadly \citep{birkstedt2023ai}. This literature sets the stage for why technical solutions are needed (e.g., an ``AI race that sacrifices safety'' \citep{Dafoe2018-fo} could imply a need for governance-oriented technical security mechanisms). More recent policy documents \citep[see e.g.,][]{oecd2019ai, g72023ai} enumerate high-level issues like privacy concerns, bias, a lack of accountability, and security in AI, while including technical, socio-political, and institutional considerations. Our work contributes by focusing more narrowly on the technical dimension of those issues. For instance, where such reports highlight AI accountability as a governance challenge, our taxonomy offers concrete technical problem areas (like verifiable auditing and data provenance tracking) that would help achieve accountability in practice.

The AI risk literature categorizes both realized harms and anticipated risks of AI \citep[see e.g.,][]{lobel2024ai, shelby2023sociotechnical, raji2022fallacy, Weidinger2022-yb, Critch2023-wp}.
For example, \citet{shelby2023sociotechnical} presented a human-centered taxonomy of sociotechnical harms from algorithmic systems, identifying five harm categories (representational, allocative, quality-of-service, interpersonal, and societal harms) based on a review of 172 research papers. Similarly, \citet{uuk2024taxonomy} conducted a systematic literature review to derive 13 categories of systemic AI risks, including environmental harms, structural discrimination, and loss of control. 
Similar to the literature on challenges in AI governance, this prior work is relevant to our efforts in that it motivates work on AI governance -- indeed, many of the open problems we list (such as developing robust evaluation methods) are motivated by those very risks.
By situating our work among these prior catalogs, we underscore that our contribution is a focused extension: we zoom in on the technical layer of AI governance to provide a structured overview of open technical problems that underlie or intersect with identified AI governance challenges.

\subsection{Scope and Limitations}\label{12-scope-and-limitations}

This paper aims to give a broad overview of open technical problems for
AI governance, identifying gaps in existing or suggested governance
proposals, while avoiding taking a normative position on their
desirability or efficacy. Indeed, the governance aims motivating some of the
open questions outlined below may be in tension with each other, and we
do not expect their solutions all to be used within the same governance
framework. For example, broad access to some AI systems may be in
conflict with ensuring their security.

At the same time, we are conscious of the potential pitfalls of
techno-solutionism -- that is, relying solely on proposed technical
fixes to complex and often normative social problems -- including a lack
of democratic oversight and introducing further problems to be fixed
\citep{Michael2020-lr,Lindgren2023-wb,Angel2024-ii,Allen2024-cd}.
Many of the TAIG tools presented below are hypothetical and speculative,
and we make no claims about the feasibility of developing solutions.
Furthermore, some of the TAIG measures highlighted are dual-use. For
example, while hardware-enabled mechanisms for monitoring advanced
compute hardware could provide increased visibility into the private
development of the largest models, they could also potentially be applied to
unreasonably surveil individuals using such hardware for legitimate
purposes.

Thus, having solutions to all open problems outlined in this paper will
not have \emph{solved} AI governance. On the contrary, careful management
will be necessary to determine a balance between capacities that are in
tension with each other, and to ensure that dual-use capacities are not
misused. Furthermore, many AI governance problems may rely predominantly
on non-technical solutions, such as ensuring the appropriate inclusion
of countries impacted by AI in international AI governance
decision-making \citep{Trager2023-qj}. However, we argue
that making progress on the technical problems outlined below can help
to ensure more robust AI governance on net.

% While related and overlapping, we view TAIG as complementary to
% sociotechnical approaches to AI safety and governance
% \citep{lazar2023ai,Bogen2024-zv,Oduro2024-zh}. In
% particular, while sociotechnical approaches view ``society and
% technology together as one coherent system''
% \citep{Chen2024-ho}, TAIG considers the instrumental value of technical work for enacting
% governance. Taken together, TAIG and sociotechnical
% approaches can serve as complementary methods for mitigating risks and
% promoting beneficial outcomes of AI
% \citep{Narayanan2023-mt}.

We view TAIG as related to (and in some places overlapping with) existing areas of research. Namely, TAIG differs from topics in AI safety and alignment in that it is not aimed at directly improving the safety of AI systems.
Furthermore, while sociotechnical approaches to AI safety and governance
\citep[see e.g.][]{lazar2023ai,Bogen2024-zv,Oduro2024-zh} view ``society and
technology together as one coherent system''
\citep{Chen2024-ho}, TAIG considers the instrumental value of technical work for enacting
governance.
Taken together with TAIG, these diverse 
approaches can serve as complementary methods for mitigating risks and
promoting beneficial outcomes of AI
\citep{Narayanan2023-mt}.
Finally, and as noted above, we view TAIG as a subset of AI governance more broadly, which also includes many topics in diverse disciplines including political science, economics, and philosophy.

We consider some notable fields, topics, and problems to be out of scope
for this paper. In particular, technical work that directly improves the
performance, safety, or robustness of AI systems, or addresses related
ethical concerns -- while highly relevant to AI governance -- is
considered out of scope. Topics regarding government or public-sector
use of AI
\citep{Margetts2019-va,Aitken2022-zg,Margetts2022-kn,Straub2023-xj}
or ways in which AI could itself be used to defend against or
ameliorate downstream harms of AI \citep{Bernardi2024-kn}
are also out of scope.

\newpage
\subsection{Reader Guide}\label{13-reader-guide}

\begin{center}
\begin{table}[h]
\centering
\caption{Relevant problem areas organized by reader background}
\label{tab:readertable}
\begin{tabular}{rll}
\toprule
%\multicolumn{2}{c}{\bfseries ML theory}\\\midrule
{\bfseries ML Theory} & Assessment & \ref{312-infrastructure-and-metadata-to-analyse-large-datasets}; \ref{313-attribution-of-model-behaviour-to-data}; \ref{321-definition-of-chip-and-cluster-specifications-for-model-training}; \ref{322-classification-of-workloads}; \ref{331-reliable-evaluations}; \ref{332-efficient-evaluations}; \ref{341-downstream-impact-evaluations} \\ 
& Access & \ref{411-privacy-preserving-third-party-access-to-datasets}; \ref{421-addressing-compute-inequities}; \ref{431-facilitation-of-third-party-access-to-models} \\
& Verification & \ref{511-verification-of-training-data}; \ref{522-verification-of-compute-workloads}; \ref{531-verification-of-model-properties}; \ref{533-proof-of-learning}; \ref{541-verifiable-audits}; \ref{542-verification-of-ai-generated-content} \\ 
& Security & \ref{611-detection-and-prevention-of-training-data-extraction}; \ref{631-prevention-of-model-theft}; \ref{632-shared-model-governance}; \ref{633-model-disgorgement-and-machine-unlearning}; \ref{641-detection-of-adversarial-attacks}; \ref{642-modification-resistant-models}; \ref{643-detection-and-authorisation-of-dual-use-capability-at-inference-time} \\
& Operationalization & \ref{702-deployment-corrections} \\\midrule
%\multicolumn{2}{c}{\bfseries Applied ML} \\\midrule
{\bfseries Applied ML} & Assessment & \ref{312-infrastructure-and-metadata-to-analyse-large-datasets}; \ref{331-reliable-evaluations}; \ref{332-efficient-evaluations}; \ref{341-downstream-impact-evaluations} \\ 
& Access & \ref{431-facilitation-of-third-party-access-to-models}; \ref{441-access-to-downstream-user-logs-and-data} \\ 
& Security & \ref{643-detection-and-authorisation-of-dual-use-capability-at-inference-time} \\ 
& Operationalization & \ref{701-translation-of-governance-goals-into-policies-and-requirements}; \ref{702-deployment-corrections} \\ 
& Ecosystem Monitoring & \ref{801-understanding-associated-risks}; \ref{802-predicting-future-developments-and-impacts}; \ref{803-assessing-environmental-impacts} \\\midrule
% \multicolumn{2}{c}{\bfseries Cybersecurity} \\\midrule
{\bfseries Cybersecurity} & Verification & \ref{522-verification-of-compute-workloads} \\ 
& Security & \ref{621-securing-ai-using-hardware-mechanisms}; \ref{623-enforcement-of-compute-usage-restrictions}; \ref{631-prevention-of-model-theft}; \ref{643-detection-and-authorisation-of-dual-use-capability-at-inference-time} \\ 
& Operationalization & \ref{702-deployment-corrections} \\\midrule
%\multicolumn{2}{c}{\bfseries Cryptography}\\\midrule
{\bfseries Cryptography} & Assessment & \ref{311-identification-of-problematic-data}; \ref{322-classification-of-workloads} \\ 
& Access & \ref{411-privacy-preserving-third-party-access-to-datasets}; \ref{412-preservation-of-evaluation-data-integrity}; \ref{421-addressing-compute-inequities}; \ref{431-facilitation-of-third-party-access-to-models}; \ref{441-access-to-downstream-user-logs-and-data} \\ 
& Verification & \ref{511-verification-of-training-data}; \ref{521-verification-of-chip-location}; \ref{522-verification-of-compute-workloads}; \ref{533-proof-of-learning}; \ref{541-verifiable-audits}; \ref{542-verification-of-ai-generated-content} \\ 
& Security & \ref{621-securing-ai-using-hardware-mechanisms}; \ref{623-enforcement-of-compute-usage-restrictions}; \ref{632-shared-model-governance}; \ref{643-detection-and-authorisation-of-dual-use-capability-at-inference-time} \\\midrule 
%\multicolumn{2}{c}{\bfseries Hardware engineering} \\\midrule
{\bfseries Hardware} & Assessment & \ref{312-infrastructure-and-metadata-to-analyse-large-datasets}; \ref{321-definition-of-chip-and-cluster-specifications-for-model-training}; \ref{322-classification-of-workloads}\\ 
{\bfseries Engineering} & Access & \ref{421-addressing-compute-inequities} \\ 
& Verification & \ref{521-verification-of-chip-location}; \ref{522-verification-of-compute-workloads} \\ 
& Security & \ref{621-securing-ai-using-hardware-mechanisms}; \ref{622-anti-tamper-hardware}; \ref{623-enforcement-of-compute-usage-restrictions}; \ref{631-prevention-of-model-theft}; \ref{632-shared-model-governance} \\\midrule
%\multicolumn{2}{c}{\bfseries Software engineering} \\\midrule
{\bfseries Software} & Assessment & \ref{311-identification-of-problematic-data}; \ref{312-infrastructure-and-metadata-to-analyse-large-datasets}; \ref{332-efficient-evaluations}; \ref{341-downstream-impact-evaluations} \\  
{\bfseries Engineering} & Access & \ref{421-addressing-compute-inequities} \\ 
& Verification & \ref{522-verification-of-compute-workloads} \\ 
& Security & \ref{621-securing-ai-using-hardware-mechanisms}; \ref{631-prevention-of-model-theft} \\\midrule
%\multicolumn{2}{c}{\bfseries Mathematics \& statistics} \\\midrule
{\bfseries Mathematics and} & Assessment & \ref{312-infrastructure-and-metadata-to-analyse-large-datasets}; \ref{341-downstream-impact-evaluations} \\ 
{\bfseries Statistics} & Ecosystem Monitoring& \ref{802-predicting-future-developments-and-impacts}; \ref{803-assessing-environmental-impacts} \\\bottomrule 
\end{tabular}
\end{table}
\end{center}

% \newpage
This paper provides a broad overview of open problems across the
taxonomy defined in Section \ref{2-taxonomy}. Given the extensive nature of the main
content (Sections \ref{3-assessment}-\ref{8-ecosystem-monitoring}), we have structured it for
selective reading:

\begin{itemize}
\item

  \textbf{Each section is self-contained}, allowing readers to focus on
  their area(s) of interest.

\item

  Each section begins with a \textbf{summary table of problem
  areas}.

\item

  \textbf{Specific open problems} within each area are highlighted in \textbf{bold}.

\item

  \textbf{Example research questions} are provided in boxes at the start
  of each subsection.

\item

  Table \ref{tab:readertable} offers suggested relevant problem areas based on \textbf{reader
  expertise}.

\item

  We attach a two-page \textbf{policy brief} in
  appendix \ref{policy-brief}.

\end{itemize}

This structure aims to facilitate quick identification of key issues and
relevant problems for readers across various backgrounds and interests.

\section{Taxonomy}\label{2-taxonomy}

The paper is organized according to a two-dimensional taxonomy, based
around capacities and targets. \emph{Capacities} encompass a
comprehensive suite of abilities and mechanisms that enable stakeholders
to understand and shape the development, deployment, and use of AI, such
as by assessing or verifying system properties. These capacities are
neither mutually exclusive nor collectively exhaustive, but they do
capture what we believe are the most important clusters of technical AI
governance. We list all considered capacities, along with descriptions,
in Table \ref{tab:overview-capacities}. An overview of our methodology, and in particular how we derived this taxonomy, can be found in Appendix~\ref{methodology}.

%%%% Table 2.1 goes here
\begin{table}[ht]
\renewcommand{\arraystretch}{1.65}
\caption{Overview of capacities and their importance for AI governance}
\centering
\label{tab:overview-capacities}
\begin{tabular}{>{\centering\arraybackslash}p{0.17\linewidth}p{0.39\linewidth}p{0.36\linewidth}}\toprule
\centering\textbf{Capacity} & \centering\textbf{Description} & \centering\arraybackslash \textbf{Why it matters for governance} \\\cmidrule(lr){1-1}\cmidrule(lr){2-2}\cmidrule(lr){3-3} 
Assessment & The ability to evaluate AI systems, involving both technical analyses and consideration of broader societal impacts. & Enables the identification and understanding of system capabilities and risks, allowing for more targeted governance intervention. \\
Access & The ability to interact with AI systems, including model internals, as well as obtain relevant data and information while avoiding unacceptable privacy costs. & Enables external research and assessment of AI systems, and aids in fairly distributing benefits of AI across society. \\
Verification & The ability of developers or third parties to verify claims made about AI systems' development, behaviors, capabilities, and safety. & Establishes trust in AI systems and confirms compliance with regulatory requirements. \\
Security & The development and implementation of measures to protect AI system components from unauthorized access, use, or tampering. & Ensures the integrity, confidentiality, and availability of AI systems and guards against misuse. \\
Operationalization & The translation of ethical principles, legal requirements, and governance objectives into concrete technical strategies, procedures, or standards. & Bridges the gap between abstract principles and practical implementation of regulatory requirements. \\
Ecosystem Monitoring & Understanding and studying the evolving landscape of AI development and application, and associated impacts. & Enables informed decision-making, anticipation of future challenges, and identification of key leverage points for effective governance interventions.\\\bottomrule
\end{tabular}
\end{table}

The second axis of our taxonomy pertains to the \emph{targets} that
encapsulate the essential building blocks and operational elements of AI
systems\footnote{(Repeat of footnote 1) Our understanding of AI systems
  follows that of \citep{Basdevant2024-qg}, encompassing
  infrastructure such as compilers, model components such as datasets,
  code, and weights, as well as UX considerations.} that governance
efforts may aim to influence or manage. They are adapted from categories
introduced in \citep{Bommasani2023-gp}. Each capacity
given above can be applied to each target. We structure our paper
around the resulting pairs of capacities and targets, with the exception of \emph{operationalization} and \emph{ecosystem monitoring} which cut across all targets. The targets
considered in this report are summarized in Table \ref{tab:overview-targets}.

%%%
%%% Table 2.2 goes here
\begin{table}[ht]
\centering
\renewcommand{\arraystretch}{1.65}
\caption{Overview of targets}
\label{tab:overview-targets}
\begin{tabular}{>{\centering}p{0.2\linewidth}p{0.56\linewidth}}\toprule  
\textbf{Target} & \centering\arraybackslash\textbf{Description}\\\cmidrule(lr){1-1}\cmidrule(lr){2-2}
Data & The pretraining, fine-tuning, retrieval, and evaluation datasets on which AI systems are trained and benchmarked. \\
Compute & Computational and hardware resources required to develop and deploy AI systems. \\
Models and Algorithms & Core components of AI systems, consisting of software for training and inference, their theoretical underpinnings, model architectures, and learned parameters. \\
Deployment & The use of AI systems in real-world settings, including user interactions, and the resulting outputs, actions, and impacts.\\\bottomrule
\end{tabular}
\end{table}

We recognize that organizational processes undertaken during the
development and deployment of AI systems intersect with and shape these
targets, and could be considered as regulatory targets in their own
right. However, we have chosen not to include them as explicit targets
in our taxonomy as processes mostly involve non-technical
challenges that fall outside the scope of our paper. In cases where
processes do face challenges, we address such issues within the context
of the most relevant target. For example, compliance with
content-creators' right to opt-out is dependent on identifying
copyrighted samples in datasets (Section \ref{31-data}). 

\newpage
\section{Assessment}\label{3-assessment}

Evaluations and assessments of the capabilities and risks of AI systems
have been proposed as a key component in AI governance regimes. For
example, \emph{model evaluations and red-teaming}\footnote{\emph{Red-teaming}
  refers to deliberately trying to find ways to make a system
  behave poorly, produce harmful outputs, or be misused, in order to
  identify potential risks and vulnerabilities to be addressed.}
comprised a key part of the voluntary commitments agreed between labs
and the UK government at the Bletchley Summit
\citep{Department_for_Science_Innovation_Technology2023-xm}.
Furthermore, the White House Executive Order on Artificial Intelligence
requires developers of the most compute-intensive models to share the
results of all red-team tests of their model with the federal government
\citep{The_White_House2023-ru}.

The purpose of assessment is to detect problematic behavior or impacts
of AI systems before resulting harms can materialize, as well as to ensure systems
are safe, robust, and non-discriminatory.
However, the assessment of some targets, especially in the context of
foundation models, is currently more an art than a science, with a
significant number of open challenges
\citep{Chang2024-jr,Weidinger2023-th}. These issues are
exacerbated by the fact that evaluations are expensive to conduct at
scale. While assessment and evaluation standards are emerging
\citep{National_Institute_of_Standards_and_Technology_NIST2023-xt,UK_AI_Safety_Institute2024-ki},
there are still fundamental open technical problems that need to be
addressed to ensure robust and informative assessments.

\begin{figure}[H]
    \centering
    \includegraphics[width=0.8\linewidth]{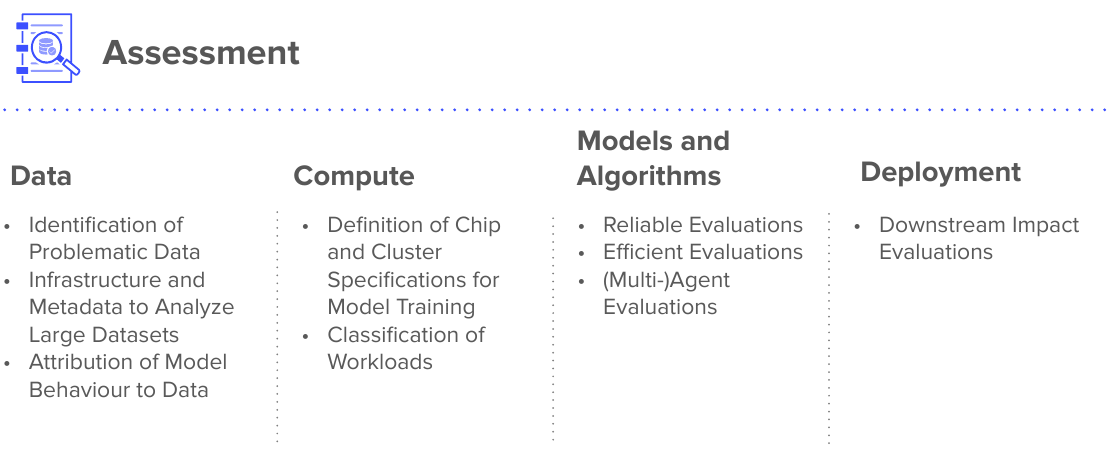}
    \caption{Open problem areas in the \emph{Assessment} capacity,
organized by target}
    \label{fig:2assessment}
\end{figure}

\subsection{Data}\label{31-data}

\begin{tcolorbox}[breakable,boxrule=1pt,enhanced jigsaw, sharp corners,pad at break*=1mm,colbacktitle=background,colback=background,colframe=black,coltitle=black,toptitle=2mm,bottomtitle=1.5mm,width=\linewidth,fonttitle=\bfseries,parbox=false,title=Example Research
Questions,boxed title style={empty,boxrule=0pt, bottom=0pt},phantom={\phantomsection\hypertarget{box1}}]

\begin{enumerate}[series=researchquestions,leftmargin=*]

\item
How can methods for identifying problematic data be scaled to large (on the magnitude of trillions of tokens/samples) datasets? (\ref{311-identification-of-problematic-data})

\item
How can license collection be automated to prevent training on unlicensed data? (\ref{311-identification-of-problematic-data})

\item
How can the accuracy of licenses be ensured when aggregating datasets from multiple sources? (\ref{311-identification-of-problematic-data})

\item
How can problematic data be identified without full/direct access to the dataset? (\ref{311-identification-of-problematic-data})

\item
How can contamination of training data with problematic samples be reliably detected? (\ref{311-identification-of-problematic-data})

\item
How can harmful data be removed from a dataset without facilitating its easy identification by malicious actors? (\ref{311-identification-of-problematic-data})

\item
What license and meta-data reporting requirements could assist in responsible data practices? (\ref{312-infrastructure-and-metadata-to-analyse-large-datasets})

\item
What infrastructure is needed to enable researchers to audit large datasets? (\ref{312-infrastructure-and-metadata-to-analyse-large-datasets})

\item
How can macro-scale dataset properties, such as persistent bias, be identified and measured? (\ref{312-infrastructure-and-metadata-to-analyse-large-datasets})

\item
What information about datasets is necessary for determining their suitability for training? (\ref{312-infrastructure-and-metadata-to-analyse-large-datasets})

\item
What is the effect of problematic data on downstream system performance? (\ref{313-attribution-of-model-behaviour-to-data})

\item
Can system behaviors and/or properties be accurately attributed to pretraining and/or fine-tuning data samples? (\ref{313-attribution-of-model-behaviour-to-data})

\end{enumerate}
\end{tcolorbox}

%%% Table 4 goes here

\subsubsection{Identification of Problematic
Data}\label{311-identification-of-problematic-data}

\uline{Motivation:} Data plays a central role in the development and
resulting capabilities of AI systems. Therefore, issues with data can
propagate downstream, resulting in undesirable properties of models. We
identify two ways in which data can be problematic.

The first is that data samples may violate some legal or ethical principle, simply by 
virtue of being included in a dataset. For instance,
the presence of a sample could constitute a copyright or privacy
violation
\citep{Brown2022-af,Rahman2023-sz,Subramani2023-rs,Marcus2024-na},
data poisoning
\citep{Biggio2012-ng,Steinhardt2017-af,Wallace2020-au,Carlini2021-oo,Carlini2021-os,Schuster2021-xt,Carlini2023-ph},
or be inherently harmful 
\citep{Thiel2023-xw,Birhane2021-de,Birhane2023-vn,Luccioni2021-bc}.

The second way that data could be problematic is if its use in training
causes undesirable downstream effects. For instance, models trained on
factually incorrect content in training data, such as vaccine
disinformation, might replicate those factual inaccuracies. Indeed,
\citep{Lin2022-pw} demonstrate how models can
``{[}generate{]} many false answers that mimic popular misconceptions.''
Alternatively, low-quality data, such as inaccuracies in low-resource
languages, can compromise performance of models in those
languages \citep{Kreutzer2022-mn}.

Being able to identify problematic data used in the development of AI systems could facilitate various governance measures. For example, developers found to be using inherently harmful data could face penalties for improper use. Alternatively, developers found to be training on data that induces harmful model behaviors could face a higher burden of proof when demonstrating that their models are sufficiently safe for public deployment.

\uline{Open problems:}

\textbf{Identifying problematic data given access to the training
data.} For model developers with full access to the dataset, the major
challenge is defining concrete, operationalizable criteria for detecting
and removing problematic samples before training. Some problematic
samples may be easier to identify than others; for instance, social
security numbers can be identified with regexes or direct copies can be
easily identified through pattern matching. However, the identification
of other forms of problematic data samples poses more of a challenge. For example,
understanding whether a data sample constitutes a copyright infringement
requires knowledge of copyright law, making judgments about how
much lexical similarity amounts to infringement, and the intended
application of the data
\citep{Henderson2023-qz,Balganesh2012-qa}. Other
approaches could resemble detection methods for data contamination, such
as fuzzy string matching, or audio and video fingerprinting, as used by
Youtube to identify copyrighted pieces
\citep{Cano2005-vm,Wu2017-kn}.

\textbf{Identifying problematic data without access to the training
data.} Regulators, auditors, or other entities who don't necessarily
have access to a system's training data may need to find proxies for
problematic data based on model behavior. Potential approaches
include calculating confidence scores for the inclusion of data points
\citep{Li2024-tf}, or using data watermarks introduced by creators
that can be detected without access to the training corpora
\citep{Wei2024-sm}. Other approaches may be used to
identify the use of copyrighted data, including inference attacks
\citep{Shokri2017-ht} and influence functions
\citep{Grosse2023-mj,Choe2024-gp}. Yet, these approaches
lack robustness -- an issue that further research could aim to address.

\textbf{Tracing data provenance.} It can be challenging for model
developers to ensure training data is correctly licensed due to licenses
frequently being aggregated and misrepresented
\citep{Longpre2023-zm}. Hence, better tooling for data
provenance will be necessary if developers are to feel comfortable that
they are honoring creators' licenses. The Data Provenance Initiative
\citep{Longpre2023-zm} conducted a large-scale audit of
open-source fine-tuning data collections and cataloged corresponding
data sources, licenses, and creators in an attempt to establish the
provenance of data.
% Model developers could use this information to
% decide which data are problematic if they don't acquire a license, and
% hence exclude that data from their training set or acquire a license.
However, the Data Provenance Initiative is limited in the datasets they
cover, and expanding it to other data is resource-intensive. Automated
license collection or standardized meta-data reporting for datasets
\citep{Longpre2024-xi} could help developers to release
systems without facing legal ambiguity.

\textbf{Silent removal of harmful data.} Another challenge is that of
being able to remove harmful data without the unwanted side-effect of
publicly identifying it, and thus allowing malicious actors an easy way
to source such data \citep{Thiel2023-xw}. For example,
the LAION-5B image dataset \citep{Schuhmann2022-gt} is
composed of image URLs and metadata, as opposed to the images
themselves. If one were to simply remove the URLs of the identified
harmful images from the dataset, 
then this could provide malicious actors with the locations of
these harmful images, either by comparing the datasets before and after
removal, or directly through a repository change log where the dataset is
hosted. However, care should be taken as methods for addressing this challenge could potentially also allow
for the subversion of existing techniques (such as open change logs) for
facilitating transparency into developers' data handling practices.

\subsubsection{\texorpdfstring{Infrastructure and Metadata to
Analyze Large Datasets
}{Infrastructure and Metadata to Analyze Large Datasets }}\label{312-infrastructure-and-metadata-to-analyse-large-datasets}

\uline{Motivation:} In addition to improved methods for auditing and
identifying problematic samples in datasets, methods and infrastructure
are needed for implementing these methods at scale. Contemporary
datasets commonly contain on the order of tens of terabytes of data,\footnote{See
  for example, FineWeb \citep{Penedo2024-js}, a dataset of
  English text totaling 45 terabytes.} introducing the
challenge of having the computational resources to store and handle
such large quantities of data. Addressing this challenge will be
crucial for enabling the identification of problematic data in practice.

\uline{Open problems:}

\textbf{Automating meta-data collection for datasets.} A challenge with
large-scale dataset infrastructure is that metadata -- including links
to the original data sources or license information -- is not always
provided \citep{Piktus2023-if}. An open research question
is how the collection of metadata 
for previously published datasets could be automated. Additionally,
exploring the automatic addition of cryptographic check sums as part of
the metadata at the dataset level could help to enable users to confirm
that the data they download matches the original data. Such an approach
could be a partial solution to ensuring that datasets have not been
altered (either maliciously or incidentally), especially in light of
advances in data poisoning attacks
\citep{Carlini2023-ph}.

\textbf{Determining relevant targets for dataset analysis.} Another
question concerns the appropriate metrics for ascertaining the
suitability of datasets for use in training, based on diffuse and
macro-scale properties. For example, a dataset may be biased not as a result of the inclusion
of individual samples, but the overall distribution of all included samples. Determining which measures and metrics are relevant in
large-scale dataset analyses is an open challenge
\citep{Cho2021-vh}. A further open question concerns the
information necessary for applying these metrics when evaluating
datasets, extending the work on model and data cards by
\citet{Mitchell2022-wb} and
\citet{Pushkarna2022-cf}.

\textbf{Developing search tools for large datasets.} While there exists
some work to quantitatively analyze dataset attributes
\citep{Mitchell2022-wb}, this methodology generally
``does not adapt well to web-scale corpora''
\citep{Piktus2023-if}. One initial attempt to close this
gap and to provide infrastructure for the quantitative and qualitative 
analysis of large datasets is the ROOTS Search Tool
\citep{Piktus2023-if}. However, at the moment, ROOTS is
limited to the 1.6TB corpus used to train BLOOM, and does not support
other large-scale datasets. Extending this tool or creating similar
tools for other open-access datasets could help with large-scale data
governance. A related effort by \citet{Elazar2024-iz} has
provided an example of such an extension.

\subsubsection{Attribution of Model Behavior to
Data}\label{313-attribution-of-model-behaviour-to-data}

\uline{Motivation:} As noted above, one way in which training data may be problematic is if
it causes undesirable downstream effects in models trained on that data,
such as reproducing false information. In order to identify samples that
should be excluded from training datasets for this reason, it may be
necessary to understand how dataset composition can affect model
performance. Thus, a complementary open issue to identifying problematic
data is attributing model behavior to specific data points.
Insights could be applied to different datasets used during development. For example, while pretraining data forms the basis for overall model performance \citep[see e.g.][]{Blakeney2024-km}, preference data used during fine-tuning has a larger impact on the extent to which the resulting model is aligned with users' interests.

\uline{Open problems:}

\textbf{Understanding how pretraining data affects model behavior.} Due
to the size and complexity of contemporary AI systems, collective 
understanding of how specific training samples
can contribute to problematic model behavior is incomplete
\citep{Lin2022-rv,Siddiqui2022-tj}.
\citet{Udandarao2024-pa} have, for example, investigated
how pretraining concept frequency impacts downstream performance. Others
have studied how upsampling domain-specific data relative to generic
web-scraped text affects model performance
\citep{Blakeney2024-km}.

\textbf{Understanding properties and effect of preference data on
fine-tuned models.} In addition to pretraining data, future work could
aim to understand the effects of the preference data used to fine-tune
models, whether this is via a reward model, constitution, or another
representation of preferences. It may be relevant to ensure that
preference data has certain properties, such as representativeness,
diversity, or neutrality, and design evaluations for the quality of
this data. Related work in this context is research on preference
aggregation for fine-tuning language models
\citep{Mostafazadeh_Davani2022-nc,Siththaranjan2024-cn,Barnett2023-bh}
and scaling laws for reward model over-optimization \citep[for example,][]{Gao2023-cw}.

\textbf{Understanding the impact of synthetic data.} Model developers
may face data scarcity as current AI systems require increasing amounts
of training data \citep{Villalobos2022-wl}. Synthetic
data generation offers an alternative to creating new authentic data,
which can be both time-consuming and resource intensive. However, the
impacts of training on synthetic data on model behavior are not yet
well understood
\citep{Guo2023-we,Alemohammad2023-jz,Martinez2023-tk,Shumailov2023-xd,Gerstgrasser2024-oo}.
Questions remain on the effect that different strategies of training on
synthetic data can have on model performance and bias, given that
synthetic data has been shown to lack representativeness and
insufficiently reflect imperfections of real-world data
\citep{Hao2024-lb}. Future research could aim to provide
greater clarity regarding the effect of training on synthetic data, or
develop uses of synthetic data for promoting model safety and reducing
bias.

\textbf{Balancing tractability and accuracy in data attribution.} One
key challenge in data attribution is the trade-off between computational
tractability and accuracy \citep[see, for example,][]{Ghorbani2019-nn,Jia2019-kl,Ilyas2022-xk,Akyurek2022-fh},
such as when using influence functions to attribute behaviors to data
examples \citep{Basu2020-nf,Grosse2023-mj,Choe2024-gp}.
\citet{Park2023-pd} aim to overcome this trade-off by
introducing TRAK (\emph{Tracing with the Randomly-projected After
Kernel}) but so far this work has only been tested for small foundation
models. Due to TRAK's methodology requiring the training of multiple
model versions for different subsets of the training set, it is unlikely
that such a method would scale to the largest models, creating an avenue
for future work.

\subsection{Compute}\label{32-compute}

\begin{researchbox}

\begin{enumerate}[resume=researchquestions,leftmargin=*]

\item

  What hardware properties or chip specifications are most indicative of
  suitability for AI training and/or inference? How does this differ
  from other scientific, business, or casual uses of high-end hardware?
  (\ref{321-definition-of-chip-and-cluster-specifications-for-model-training})

\item

  How efficiently can AI models be trained using a large number of small
  compute clusters? (\ref{321-definition-of-chip-and-cluster-specifications-for-model-training})

\item

  How can decentralized training attempts be identified? (\ref{321-definition-of-chip-and-cluster-specifications-for-model-training})

\item

  Can large training runs be detected while retaining developer privacy,
  for example through identifying signatures in processor utilization?
  (\ref{322-classification-of-workloads})

\item

  Can compute workloads be reliably classified as either training,
  inference, or non-AI-related, for example through identifying
  signatures in processor utilization? (\ref{322-classification-of-workloads})

\end{enumerate}
\end{researchbox}

\subsubsection{Definition of Chip and Cluster Specifications for
Model
Training}\label{321-definition-of-chip-and-cluster-specifications-for-model-training}

\uline{Motivation:} Compute governance has been proposed as a lever for
governing advanced AI systems, due to the large amounts of computing
resources required for their training and deployment
\citep{Sastry2024-sf}. However, despite its potential
efficacy, compute governance is a blunt instrument, with previous actions including the
restriction of the sale of a broad range of high-end chips to China
\citep{Bureau_of_Industry_and_Security2022-md,Bureau_of_Industry_and_Security2024-xy}.
It would thus be beneficial to be able to limit compute governance
interventions to only the chips or compute clusters that are most
relevant for developing and deploying AI systems of interest to
policymakers. Ensuring that compute governance is targeted only where
needed will require thoughtful derivation of metrics and specifications
that capture hardware of concern, while excluding the vast majority of
computing resources that are not used for industrial-scale AI.

\uline{Open Problems:}

\textbf{Assessing the effect of chip specifications on AI workload
suitability.} One open issue is understanding how different chip
specifications -- such as throughput, memory bandwidth, memory capacity,
and interconnect bandwidth -- affect a chip's
suitability for different AI workloads, or the suitability of a cluster
made up of those chips. Previous regulations concerning chips have arguably
contained loopholes due to limitations in their technical
specifications. For example, NVIDIA's A800, ``while
compliant with October 7 controls, is still capable of performing
complex AI tasks, and was a top seller in China before the updated
October 17, 2023 controls.'' \citep{Reinsch2023-ue}

\textbf{Understanding the implications of decentralized training and
cluster size.} A related sub problem is understanding decentralized
training, specifically the question of how efficiently AI models can be
trained using multiple geographically disparate compute clusters.
There is substantial technical work on developing
decentralized training methods \citep[see, for example,][]{Douillard2024-bp}. Another open problem is the
efficiency and cost impact of using a larger number of less powerful
chips within a cluster, as opposed to using a smaller number of more
powerful chips totaling the same theoretical throughput, sometimes
known as \emph{slicing}.

\subsubsection{Classification of Workloads}\label{322-classification-of-workloads}

\uline{Motivation:} In addition to classifying hardware, it is also
useful to classify computational workloads in order to identify
potentially concerning or anomalous workloads. For example, Executive Order 14110
requires the reporting of training runs above a particular compute
threshold \citep{The_White_House2023-ru}. Being able to classify workloads while preserving customer
privacy could assist with such reporting, as well as a range of other
governance goals, such as identifying compute usage trends, and audit trails
for the development of powerful models
\citep{Heim2024-cj,Shavit2023-wc}.

\newpage
\uline{Open Problems:}

\textbf{Privacy-preserving workload classification.} Compute providers,
such as data center operators and cloud computing firms, typically
already collect a wide variety of high-level data on customers and
workloads \citep{Heim2024-cj}. An open question is thus whether it
is possible to use this data to develop reliable workload classification
techniques, for example, determining whether a training workload exceeds
certain compute thresholds, or whether an inference workload involves
malicious cyberactivity
\citep{Commerce_Department2024-jw}. Such techniques
would need to account for changes in the hardware, software packages,
and specific algorithms used in AI workloads over time.

\textbf{Ensuring workload classification techniques are robust to
adversarial gaming}. Adversarial compute customers may try to obfuscate
their activities to avoid workload classification, by introducing noise
in the way computational resources are used, or breaking up workloads
across multiple cloud accounts, providers, or computing clusters.
Designing workload classification approaches that are robust to this
kind of gaming, or otherwise are able to detect when this kind of gaming
is occurring, is an open challenge
\citep{Egan2023-dx,Heim2024-cj}.

\subsection{Models and Algorithms}\label{33-models}

\begin{researchbox}

\begin{enumerate}[resume=researchquestions,leftmargin=*]

\item

  How can the thoroughness of evaluations be measured? 
  (\ref{331-reliable-evaluations})

\item 
  
  How can potential
  blind spots of evaluations be identified? (\ref{331-reliable-evaluations})

\item

  How can potential data contamination be accounted for when conducting
  evaluations? (\ref{331-reliable-evaluations})

\item

  How can mechanistic analysis of model internals, such as weights,
  activations and loss landscapes on particular data, be used to improve
  understanding of a model's capabilities, limitations and weaknesses? (\ref{331-reliable-evaluations})

\item 
  
  How generalizable are mechanistic analyses across models? (\ref{331-reliable-evaluations})

\item

  How can methods for red-teaming models be scaled and/or automated?
  (\ref{332-efficient-evaluations})

\item

  How can the capabilities and risks of AI agents be evaluated? (\ref{333-multi-agent-evaluations})

\item

  How can the capabilities and risks of networks of multiple interacting
  AI agents be evaluated? (\ref{333-multi-agent-evaluations})

\end{enumerate}
\end{researchbox}

\subsubsection{Reliable Evaluations}\label{331-reliable-evaluations}

\uline{Motivation:} A great deal of research is focused on evaluating AI
models to measure their performance and identify capabilities and
failure modes, including by state actors
\citep{Department_for_Science_Innovation_Technology_undated-eg,UK_AI_Safety_Institute2024-ki,Anthropic2024-rj}.
Yet, state-of-the-art AI systems can exhibit unpredicted downstream
capabilities that often evade evaluations
\citep{Shayegani2023-cd,Carlini2023-na,Schaeffer2024-vb}.
The rapid advancement and widespread deployment of AI systems has
prompted major regions, including the European Union, China, and the
United States, to put forward requirements for evaluating, reporting,
and mitigating risks associated with these systems
\citep{Reuel2024-ag}. For instance, Article 55 of the EU AI Act
mandates that providers of general-purpose AI models with systemic risk
perform model evaluations using standardized protocols and tools,
including adversarial testing, to identify and mitigate systemic risks
\citep{Council_of_the_European_Union2024-pt}. Despite jurisdictions mandating capability
evaluations across various risk areas, there is a lack of technical
clarity on how to perform these assessments comprehensively and reliably
\citep{Chang2024-jr,Zhou2023-ec}, and for some risks such
evaluations simply do not yet exist
\citep{Weidinger2023-th}. In addition, evaluations for
decision-making systems in high-stakes settings will likely demand a
higher level of confidence than other applications, but it is unclear
how to determine the required level of rigor based on use case.

\uline{Open Problems:}

\textbf{Ensuring sufficient testing.} Determining whether an evaluation
procedure has identified all, if not most, of the vulnerabilities of a
system is an open problem. This is especially relevant if the
evaluation pertains to capabilities, such as deception and long-horizon
planning, that could enable harmful forms of misuse or make systems hard
to oversee or control
\citep{Park2024-pc,Shevlane2023-ov,Hendrycks2023-oz,Kinniment2023-ix,Li2024-nw,Phuong2024-qu,Bengio2024-hs}.
This issue is most acute in the case behavioral evaluations that
directly test models' performance on benchmarks or test cases
\citep{Liang2023-gh,Srivastava2023-vk,Gao2023-rc,Wang2023-qa,Lee2023-zz,Biderman2024-ch},
as such evaluations can fail to fully inform evaluators about a model's
capabilities, with particular uncertainty around capabilities that the
model may lack \citep{Casper2024-lv}. Specifically, if a
model does not exhibit a particular behavior during testing, it is
challenging to determine whether this is due to the model genuinely
lacking the underlying ability or if the evaluation method applied was
insufficient to surface it \citep{Wei2022-uy,Zhu2023-vz},
for example, due to prompt sensitivity
\citep{Zhu2023-vz,Sclar2023-no}. This ambiguity can lead
to an incomplete understanding of a model's true
capabilities and potential risks
\citep{Barrett2023-kg,Schaeffer2023-ee,Raji2021-ku}, and
means that behavioral evaluations can only offer a lower bound on a
model's potential to exhibit harmful behavior
\citep{Goel2024-ms,Casper2024-wv}. This motivates
research into how to expand the scope of behavioral evaluations as well
as how to estimate the robustness and extensiveness of existing
evaluations \citep{Chan2024-mv}.\footnote{Beyond
  insufficient model-level testing, evaluations along the whole AI
  life cycle are lacking. We address this problem separately in Section
  3.4.}

\textbf{Improving evaluations using mechanistic analysis.} The
shortcomings of behavioral evaluation techniques motivate additional
work on evaluation methods that do not suffer from the same limitations.
One proposed approach toward this is to study their internal mechanisms
\citep{Burns2022-sp,Olah2023-go,Carranza2023-tz,Casper2024-lv,Bereska2024-in}.
Although evaluations that involve developing interpretations of a
model's internal structure have seen a great deal of interest, they
remain largely untested in practice \citep[although see][]{Templeton2024-ni,Gao2024-lh}. Furthermore,
interpreting the internal mechanisms of models inevitably involves
reductions in the complexity of the system being studied
\citep{Gilpin2018-oi}.

\textbf{Testing the validity of evaluations.} It can be hard to have
high confidence that the results of the evaluations reflect properties
of the model rather than the evaluation methodology employed -- that is,
that the evaluation is internally valid.\footnote{Internal validity
  refers to the extent to which an evaluation accurately measures what
  it intends to measure within its specific context, while external
  validity refers to how well the evaluation results can be generalized
  or applied to other situations, populations, or contexts beyond the
  original study.} Making progress on this could be challenging due to
uncertainties pertaining to both evaluations and the models to which
they are applied, making it difficult to attribute results to either the
evaluation or the model. One potential avenue for progress could be
using \emph{model organisms} -- smaller simpler AI models that have
particular properties by construction \citep[for example][]{Hubinger2024-bw} -- to test whether evaluations
for the constructed property are able to reliably detect it. However, this method
would likely not be of use if trying to develop evaluations for model
properties that are only exhibited by the most capable models. More
generally, designing meta-evaluations that assess the reliability and
consistency of evaluation methodologies across different models and
contexts remains an open research challenge.

\textbf{Establishing a causal relationship between procedural design
choices and system characteristics.} Similar to attributing a model's
properties to characteristics of its training data (see Section \ref{313-attribution-of-model-behaviour-to-data}),
it may be possible to attribute behavior or performance to design
decisions made during a model's development process \citep[see, for example,][]{Simson2024-dg}. Establishing causal relationships
between such decisions and resulting system properties may allow for
greater standardization of development best practices.

\textbf{Understanding potential risks and capabilities of future AI
systems.} Finally, it is also difficult to study certain risk scenarios
that might emerge through advanced capabilities -- for example, the
ability to plan over a long horizon -- due to their hypothetical nature.
It could be useful to develop and study specific demonstrations of
harmful behavior and the effectiveness of current safety techniques
against these behaviors in current AI systems, similar to the model
organisms approach suggested in the previous paragraph
\citep{Scheurer2024-lp,Jarviniemi2024-cs,Hubinger2024-bw}.

\subsubsection{Efficient
Evaluations}\label{332-efficient-evaluations}

\uline{Motivation:} Ideally, it would be possible to test an AI system
under all possible inputs to ensure that it would not produce a harmful
output for any of them. However, performing a brute-force search over
possible inputs is intractable due to the astronomically large input
spaces for modern systems,\footnote{For example, there are vastly more
  possible $20$-token strings of text or $10\times 10$ pixel images than there are
  particles in the observable universe. (One current estimate for the
  number of particles in the observable universe is
  $10^{80}$. GPT-2's tokenizer had a vocabulary size of
  $50,257$, meaning that there are approximately
  $50,000^{20} \approx 10^{94}$ unique strings of
  20 tokens. There are $256^{100} \approx
  10^{240}$ possible $10\times 10$ gray-scale images with integer pixel values in the range $[0,255]$.)} and the fact that whether an
output is harmful may be unclear or context-dependent. As a result,
current evaluation methods manually apply heuristics to guide
vulnerability searches towards regions of the input space assumed to be
more concerning -- as observed in voluntary audits by developers
\citep{OpenAI2024-wm,Kinniment2023-ix,Touvron2023-nr,Anthropic2023-mh}.
However, manual attacks quickly become impractical, expensive, and
insufficient for conducting scalable evaluations
\citep{Ganguli2022-ql}, especially when searching across
modalities and languages \citep{Ustun2024-el}. This
problem is exacerbated for increasingly capable, general-purpose systems
that have a significantly larger attack surface than narrower systems.
The development of more efficient automated approaches for identifying
model vulnerabilities will be crucial if results from model evaluations
are to be applied as inputs to governance-relevant decisions.

\uline{Open Problems:}

\textbf{Making comprehensive red-teaming less resource-intensive.}
Recent progress has been made on automated red-teaming by using
generative AI models to produce test cases
\citep{Perez2023-sl,Shah2023-ma}, develop adversarial
prompts
\citep{Deng2022-vv,Perez2022-eu,Mehrabi2023-dr,Hubinger2024-bw,Casper2023-mv,Hong2024-kr},
and automatically evaluate the outputs of other models
\citep{Zheng2023-wz,Chiang2023-rz,Ye2023-sc,Kim2023-uw,Chao2024-rz,Souly2024-zk}.
Furthermore, automated search methods have been applied to find
adversarial attacks as a way of enhancing or replacing manual methods
\citep{Wallace2019-is,Song2020-zh,Shin2020-rf,Guo2021-vm,Shi2022-rt,Kumar2022-ot,Wen2023-pe,Jones2023-lh,Zou2023-bz,Liu2023-ib,Zhu2023-cj,Andriushchenko2024-dx}.
However, despite these approaches for automating evaluations, thorough
red-teaming remains a labor-intensive process, and many failure modes
still evade red-teaming efforts
\citep{Shayegani2023-cd,Carlini2023-na,Longpre2024-zo}.
These challenges in part stem from how existing automated techniques are
often computationally expensive and crude, requiring a large degree of
guidance from human engineers \citep{Mazeika2024-fl}.
Qualitatively different approaches that automate some or all the
red-teaming process -- perhaps through the use of agentic AI systems
that can plan, use tools, and dynamically evaluate systems -- could also
allow for more scalable evaluation.

\subsubsection{(Multi-)Agent
Evaluations}\label{333-multi-agent-evaluations}

\uline{Motivation:} \emph{Agentic} AI systems are generally
characterized by an ability to accomplish tasks from high-level
specifications, directly influence the world, take goal-directed
actions, and perform long-term planning
\citep{Chan2023-aj,Durante2024-ah,Huang2024-xw}. These
capabilities could allow agentic systems to perform tasks with little
human involvement and control. While economically useful, for example,
as customized, personal assistants, or for autonomously managing complex
supply chains, agentic systems could pose unique risks due to their
ability to directly act in the world, potentially with
difficult-to-predict impacts
\citep{Chan2023-aj,Lazar2024-wz,Gabriel2024-up,Bengio2024-hs}.

\uline{Open Problems:}

\textbf{Evaluating and monitoring agentic systems.} User
customizability, such as through prompting or the integration of new
tools, makes it particularly difficult to foresee the use cases and
potential risks of agents
\citep{Shavit2023-hi,Kolt2024-cm,Cohen2024-jl},
motivating potential measures for tracking and monitoring their actions
\citep{Chan2024-ox}. Furthermore, evaluating agents is a
nascent field with significant challenges -- existing agent benchmarks
often don't have adequate holdout datasets, causing existing agents to
game and overfit to the benchmark, which in turn results in unreliable
evaluations of these systems \citep{Kapoor2024-sn}.
Similar to non-agent benchmarks (see above), best practices are
currently lacking, leading to inconsistencies across evaluations and
limiting their reproducibility \citep{Kapoor2024-sn}.
Thus, future work could aim to introduce best practices for evaluating
agentic systems.

\textbf{Expanding limited multi-agent evaluations.} On top of the
difficulties of studying single-agent systems, multi-agent interactions
add an additional layer of complexity due to information asymmetries,
destabilizing dynamics, and difficulties in forming trust and
establishing security. These problems can lead to unique complexity and
failure modes \citep{Hammond24-ft,Chan2023-bf,Akata2023-cz,Mukobi2023-eh}. In
addition, it may be the case that collectives of agents exhibit unpredictable
capabilities or goals not attributable to any one agent in
isolation \citep{Hammond24-ft}. If AI agents become increasingly
embedded in real-world services, such as in finance or the use of web
services, it will be relevant to understand such multi-agent dynamics.

\textbf{Attributing downstream impact to individual agents.} For issues
of liability, it will be critical to be able to determine which agent(s)
or system(s) can be held responsible for a particular decision or
action, if any. This may be complicated for cases in which the cause is
not solely attributable to a single AI agent.\footnote{See, for example,
  \citep{Wex_Definitions_Team2023-al}} Having methods
of tracing multi-agent interactions and determining the cause of a
particular outcome could help to solve this problem. An open technical
question in this context regards techniques for monitoring individual
agents' contributions to multi-agent systems, in order to ease
attribution of responsibility \citep{Friedenberg2019-wk}.

\subsection{Deployment}\label{34-deployment}

\begin{researchbox}

\begin{enumerate}[resume=researchquestions,leftmargin=*]

\item

  How can the downstream societal impacts of AI systems be predicted
  and/or determined? (\ref{341-downstream-impact-evaluations})

\item

  How can downstream impact evaluations be scaled across languages and
  modalities? (\ref{341-downstream-impact-evaluations})

\item

  How can benchmarks be designed in a way that ensures construct
  validity and/or ecological validity? (\ref{341-downstream-impact-evaluations})

\item

  How can dynamic simulation environments be designed to better reflect
  real-world environments? (\ref{341-downstream-impact-evaluations})

\end{enumerate}
\end{researchbox}

\subsubsection{Downstream Impact
Evaluations}\label{341-downstream-impact-evaluations}

\uline{Motivation:} The performance of models in
isolation is an imperfect proxy for the impact that AI systems will have 
in everyday use. Thus, comprehensively understanding the impacts that AI
could have on society demands robust methods for evaluating systems in
dynamic, real-world settings \citep{Ibrahim2024-vh}.
Having such methods would allow policymakers a higher-fidelity picture
of where governance intervention might be necessary in order to address potential
harms.

\uline{Open Problems:}

\textbf{Predicting the downstream societal impacts of AI systems.}
Although understanding overall societal impact is an overarching goal of
much work on evaluating AI, it is a difficult, large-scale
sociotechnical problem
\citep{Dolata2022-ws,Solaiman2023-yt,Rakova2023-et,Weidinger2023-th,Dobbe2024-uf,Bengio2024-hs}.
For example, simplified technical proxies for complex concepts like
\emph{fairness} and \emph{equity} insufficiently measure the disparate impact of
AI systems on diverse communities
\citep{Blodgett2020-ga,Selbst2021-es}. It may also be
logistically challenging to conduct downstream evaluations. For example,
while it is known that large language models exhibit significant
cross-lingual differences in safety and capabilities
\citep{Yong2023-vp,Wang2023-gp,Wang2023-xr,Jin2023-yn,Ustun2024-el},
it is expensive and time-consuming to thoroughly evaluate the
cross-lingual properties of models due to the required coordination
between speakers of many languages. Ultimately, thoroughly assessing
downstream societal impacts requires nuanced analysis,
interdisciplinarity, and inclusion
\citep{Hagerty2019-zc,Bengio2024-hs}. While recent work
has taxonomized societal impacts and provided an overview of early techniques for
their evaluation
\citep{Moss2021-sx,Shelby2022-iw,Raghavan2023-jb,Solaiman2023-yt,Weidinger2023-th,Weidinger2024-dg},
there remains a lack of structured, effective methods to quantify and
analyze these impacts.

\textbf{Ensuring construct validity of evaluations.} It can be difficult
to establish confidence that the proxy used in an evaluation or
benchmark accurately captures the concept it aims to measure -- that is,
its construct validity
\citep{Raji2021-ku,Bowman2021-ld,Hutchinson2022-qb,Subramonian2023-hm,McIntosh2024-ap}.
For example, while MMLU \citep{Hendrycks2021-ea} claims
to assess a model's understanding and memorization of knowledge from pretraining through the proxy of performance on question answering, it is unclear how well the ability to accurately
answer questions serves as an indicator for understanding. Future
research could aim to evaluate the construct validity of current
benchmarks as well as ensure construct validity for AI system
evaluations, perhaps taking inspiration from prior work in psychology
(see, for example,
\citep{Westen2003-um,Strauss2009-eo,Smith2005-pe}).

\textbf{Ensuring} \textbf{ecological validity of evaluations.} In
addition to construct validity, establishing the ecological validity of
benchmarks is an open challenge.\footnote{\emph{Ecological validity}
  refers to the extent to which results from experiments generalize to
  contexts outside of the testing environment. Compare this to
  \emph{construct validity} which refers to the extent to which an
  assessment measures the target construct.} Current benchmarks tend to
be biased towards easily-quantifiable and model-only metrics,
potentially making them ill-suited for predicting how well models
perform when deployed in real-world settings
\citep{Ouyang2023-ve,Lee2023-os}. Future work could aim
to assess the correlation between performance on benchmarks and
downstream performance, as well as propose evaluation methods for which
this correlation is tighter. Another consideration is that benchmarks
used for assessing capabilities are oftentimes used to guide development
of models, limiting the extent to which such benchmarks can be seen as
unbiased
\citep{Marie2021-ob,Dehghani2021-pz,Madaan2024-qt,Salaudeen2024-mu},
and making it challenging to know how to interpret resulting scores.

\textbf{Designing dynamic evaluations and real-world simulation
environments.} Despite many user interactions with AI systems taking
place in dynamic, multi-turn environments, current benchmarks only
evaluate performance in such settings to a limited degree
\citep{Ibrahim2024-vh}. By creating dynamic evaluation
frameworks \citep[for example, ][]{Dynabench2023-dz,Park2023-yg}, researchers could better assess a
system's performance, inherent characteristics, and potential risks in a
more realistic manner compared to static test sets. This would require
significant investment in infrastructure and tooling, such as
sophisticated simulated environments tailored to specific domains like
hacking, persuasion, or biosecurity. Similarly, experiments with human
subjects are valuable to understand the risks from human capability
increases through AI models \citep{Mouton2023-tt}, but
are less common due to their resource-intensiveness.

\newpage
\section{Access}\label{4-access}

\vspace{-0.5cm}

Many governance actions will likely require third-party access to system
components, along with the provision of resources such as compute. As
examples, external access will likely be a necessary consideration for
facilitating third-party audits
\citep{Raji2022-am,Anderljung2023-sb,US_National_Telecommunications_and_Information_Administration2024-ib,Mokander2023-hj,Casper2024-lv},
evaluations by government AI safety institutes
\citep{Govuk2023-qx,National_Institute_of_Standards_and_Technology_NIST2024-pz,Department_for_Science_Innovation_and_Technology2023-tw},
and independent academic research
\citep{Bucknall2023-fi,Kapoor2024-ss,Longpre2024-zo,House_of_Commons_Science_Innovation_and_Technology_Committee2024-gy}.

There are numerous reasons for why it may be desirable to enable these
functions to be performed by those other than the system developers.
Firstly, actions such as evaluation and auditing could benefit from the
independence of being conducted by parties external to AI developers, so
that developers do not ``mark {[}grade{]} their own homework''
\citep{Gerken2023-ck}. Furthermore, AI developers may not
have the capacity or incentives to conduct research to the extent needed
for advancing our scientific understanding of AI systems at a rate
comparable to that of advances in AI development, motivating the need
for involvement of the broader academic community. Third-party access
may also allow for broader cultural diversity and representation in the
development and governance of AI
\citep{Dobbe2020-pc,Delgado2023-cb,Held2023-en,Crowell2023-pm},
though should not be seen as a sufficient measure
\citep{Chan2021-px,Sloane2022-ze}.

However, many of the aforementioned actions are precluded due to
insufficient external access to relevant system components, especially
in the case of the most state-of-the-art systems
\citep{Solaiman2023-es,Bommasani2023-pa,Bucknall2023-fi,Casper2024-lv}.
This is often due to concerns regarding developers' intellectual
property, privacy of data subjects, legal uncertainty, and the safety of
the system in question \citep{Seger2023-lz}.

Access to compute and other resources peripheral to the systems under
consideration, such as training data, is also essential for many
auditing and research functions
\citep{Ahmed2020-li,Besiroglu2024-fx,Ojewale2024-vr}. The
challenges associated with facilitating access to such resources will
also need to be addressed if these functions are to be fulfilled in
academia and the public sector.

%\vspace{-1.5cm}

\begin{figure}[h]
    \centering
    \includegraphics[width=0.8\linewidth]{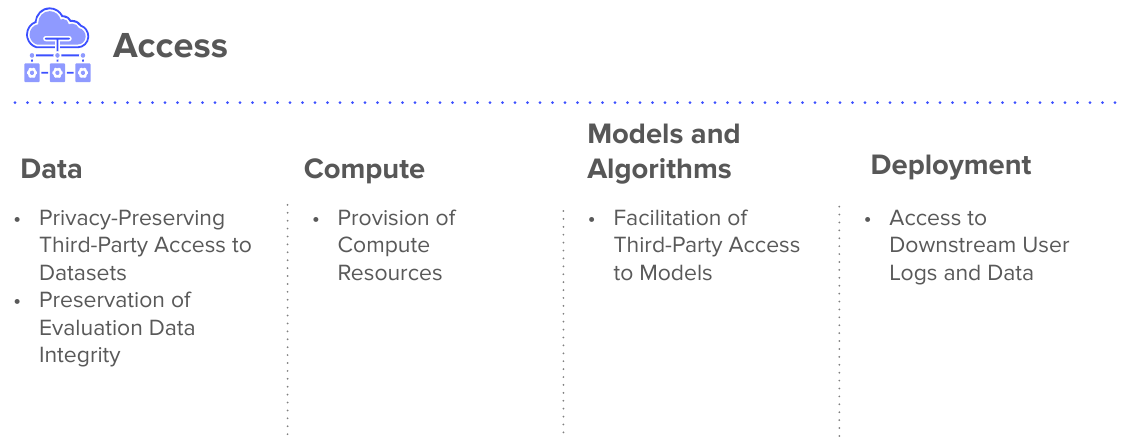}
    \caption{Open problem areas in the \emph{Access} capacity, organized by
target}
    \label{fig:3access}
\end{figure}

\subsection{Data}\label{41-data-1}

\begin{researchbox}

\begin{enumerate}[resume=researchquestions,leftmargin=*]

\item

  How can data access be structured so as to preserve privacy while
  enabling meaningful auditing? (\ref{411-privacy-preserving-third-party-access-to-datasets})

\item

  How can data access be reconciled with privacy-preserving machine
  learning? (\ref{411-privacy-preserving-third-party-access-to-datasets})

\item

  How can openly hosted datasets be prevented from contaminating
  training data? (\ref{412-preservation-of-evaluation-data-integrity})

\item

  How can independent evaluation on standardized datasets be facilitated
  without openly hosting evaluation datasets? (\ref{412-preservation-of-evaluation-data-integrity})

\end{enumerate}
\end{researchbox}

\subsubsection{Privacy-Preserving Third-Party Access to
Datasets}\label{411-privacy-preserving-third-party-access-to-datasets}

\uline{Motivation:} Access to training datasets is crucial for enabling
external data audits that aim to identify instances of harmful,
personal, or inappropriate data being included in datasets
\citep[see also Sections \ref{31-data} and \ref{51-data-2}]{Thiel2023-xw,Birhane2021-de,Birhane2023-vn,Subramani2023-rs,Luccioni2021-bc}. External access to datasets could also
be instrumentally necessary for facilitating assessment and research
into models \citep{Bucknall2023-fi,Ojewale2024-vr}, for
example, as research into how the content of training datasets influence
downstream model behavior depends on visibility into the training
datasets \citep[for example][see also Section
\ref{31-data}]{Udandarao2024-pa}.

However, na\"ively providing unrestricted access to datasets may violate
legal bounds, for example by reproducing copyrighted data, or otherwise
raise security concerns. Additional challenges stem from the lack of
legal clarity regarding how to define problematic data, such as what
constitutes a copyright violation in the context of AI training
\citep{Henderson2023-qz,Quang2021-di}. Furthermore,
developers may be reluctant to provide unrestricted access to
proprietary datasets due to the high costs associated with collating
such datasets,\footnote{Including the cost of data collection and the
  resources to ensure that licenses are correctly represented to avoid
  incorrect infringements.} the risk of leaking sensitive intellectual
property, or, in some cases, risking disclosing having knowingly trained
on illegally collected data. These concerns may be able to be alleviated
by providing greater transparency into the makeup, contents, and
aggregate characteristics of datasets, as well as information about
sources and compilation methodologies, in the absence of unrestricted
access to the entire dataset.

\uline{Open problems:}

\textbf{Structuring external access to datasets to reduce privacy
risks.} Research could aim to develop methods for providing sufficiently
deep access to datasets, for example, for the purpose of auditing and
evaluation, while protecting privacy of data subjects. Existing work in
this direction includes \citet{Trask2023-ik} which
suggests a framework for allowing third-parties to propose and execute
approved queries on AI systems and third-party data without exposing
sensitive information beyond the explicitly authorized results.
Similarly, \emph{Project Oak} is a software package for providing
``security and {[}...{]} transparency around how data is used and
by whom, even in a highly distributed system'', through the use of
trusted execution environments (TEEs) on specialized hardware
\citep[see also Section \ref{62-compute-3}]{noauthor_2024-bo}.
Inspiration can also be taken from other industries, including
healthcare \citep{noauthor_2024-bu}. For example,
\emph{OpenSAFELY} provides a platform for the analysis of patient
healthcare data for the purposes of academic research through a
combination of pseudonymization, methods for working with data \emph{in
situ} while obfuscating raw patient data, and providing transparency
into researchers' use of the platform
\citep{Bennett_Institute_for_Applied_Data_Science2024-jt}.
Future research could draw insight from this case to propose methods for
safely accessing user data for research into the societal impacts of AI.

\textbf{Addressing the tension between data access and
privacy-preserving machine learning.} A significant body of research has
explored how we can train machine learning models when data-owners do
not trust model-trainers -- for example through federated learning
\citep{Kairouz2019-iy} and training on encrypted data
\citep{Xie2014-tc,Nandakumar2019-ml}. In such cases, it
can be challenging to provide data access given that the developer
themselves lacks such visibility. Future work could explicitly address
this tension, proposing methods that allow for the auditing of
data that may have been encrypted during training.
% For example, it may
% be possible to evaluate some properties by computing a function of the
% encrypted data, just as in the case of training on encrypted data.

\subsubsection{Preservation of Evaluation Data
Integrity}\label{412-preservation-of-evaluation-data-integrity}

\uline{Motivation:} Current standardized methods for evaluating models
often utilize openly-available datasets, with the goal of comparing
model performance like-for-like (Reuel et al., \emph{forthcoming}). At present, this
is largely achieved by hosting such datasets openly in online
repositories such as \emph{HuggingFace}\footnote{\href{https://huggingface.co/}{https://huggingface.co/}}
\citep[see, for example,][]{Hendrycks2021-ea,Srivastava2023-vk,Gao2023-rc}.
However, openly hosting evaluation datasets online risks their inclusion 
in web-scraped training datasets
\citep{Deng2023-be}, either accidentally, or intentionally as a method for artificially inflating benchmarking results. Such contamination of
training data has serious implications for the efficacy and reliability
of these standardized metrics
\citep{Oren2023-qg,Roberts2023-zk,Jiang2024-xp,Zhang2024-hz,Schaeffer2023-pa}%,Susan_Zhang_suchenzang2023-wn}.

\uline{Open problems:}

\textbf{Identifying and mitigating contamination of training datasets.}
Current approaches to mitigating contamination of training datasets are
rudimentary. A potential post-contamination approach is to detect data
contamination \citep{Dong2024-dh,Golchin2023-vy} and then
correct for it when scoring benchmarks. For example, OpenAI and Meta
measured which benchmarks' test samples were potentially included in the
pretraining data of GPT-4 and Llama 2, respectively, and reported how
scores differed between contaminated test samples and non-contaminated
test samples \citep{OpenAI2024-wm,Touvron2023-nr}.
\citet{Zhang2024-hz} attempted to correct for data
contamination by creating a version of the benchmark GSM8k that is
comparable in terms of tasks, complexity and human solve rates, and
reported which models' scores dropped precipitously. Other approaches
pertain to the design of benchmarks that are robust against contaminated
models by using templates from which variations of a task can be
generated \citep{yu2023skillmix,srivastava2024functional} or by being
frequently updated \citep{white2024live}. However, the frequent updating approach is resource
intensive, especially if covering a large span of test tasks and fields, and
designing templates to support variations of tasks may
not be feasible for tasks that don't follow a predictable structure.
Alternatively, \citep{Srivastava2023-vk} make use of a
\emph{canary string}, that is, a globally unique identifier 
that is included in all sub repositories of the BIG-bench collection in
order to ease identification of these test samples in training datasets.
BIG-bench also includes a dedicated \texttt{training\_on\_test\_set} task,
which serves as a ``post-hoc diagnosis of whether BIG-bench data
was used in model training'' \citep{Srivastava2023-vk}.
However, use of canary strings depends on any and all copies of the
repositories to also include the string, and thus is not robust to
negligent users.

\textbf{Evaluating on private or encrypted evaluation datasets.}
Alternatively, research could aim to develop ways in which models can be
independently evaluated on a private or encrypted test set. Indeed, some
recent benchmarking datasets have only been made available only through
a custom evaluation API \citep[see, for example,][]{Sawada2023-el}. Furthermore, popular dataset repositories including
\emph{HuggingFace}, as well as competition platforms such as
\emph{Kaggle}, gate access to evaluation datasets in order to reduce the
risk of contaminating training data.\footnote{\citep{Hugging_Face2024-ke,Kaggle2024-cc}}
Finally, \citet{Bricman2023-bx} proposes
\emph{hashmarks}, a ``protocol for evaluating language models in
the open without having to disclose the correct answers'' by
cryptographically hashing a benchmark's reference solutions before
publication. Further work could develop this, and similar protocols for
reliably evaluating system capabilities on private evaluation data.

\subsection{Compute}\label{42-compute-1}

\begin{researchbox}

\begin{enumerate}[resume=researchquestions,leftmargin=*]

\item

  How can public compute resources be allocated fairly and equitably
  between users? (\ref{421-addressing-compute-inequities})

\item

  How can public compute infrastructure be developed in a way that
  ensures interoperability between models and software packages? (\ref{421-addressing-compute-inequities})

\item

  How can assurance be given that researcher compute provisions are
  being used for intended and stated purposes? (\ref{421-addressing-compute-inequities})

\end{enumerate}
\end{researchbox}

\subsubsection{Addressing Compute Inequities}\label{421-addressing-compute-inequities}

\uline{Motivation:} Compute usage by private companies in training and
running models has increased exponentially in the past years, and now
greatly exceeds the compute resources available for non-industry
researchers \citep{Maslej2024-tw,Besiroglu2024-fx}. While
some researchers have found that the majority of academic researchers do
not feel primarily constrained by compute access
\citep{Musser2023-gg}, others have found that this access
inequality is specifically limiting researchers'
contribution to frontier research
\citep{Ahmed2020-li,Besiroglu2024-fx,Birhane2023-vn}. To
address these concerns, there have been proposals and funding for public
compute infrastructure
\citep{noauthor_2023-cq,Ho2021-ll,Organisation_for_Economic_Co-Operation_and_Development2023-xg,National_Artificial_Intelligence_Research_Resource_Task_Force2023-mz,UK_Research_and_Innovation2023-rw}.
Though the success of these initiatives primarily depends on raising
funds to purchase sufficient compute, technical advances could still be
instrumental in their success.

\uline{Open problems:}

\textbf{Ensuring interoperability of public compute resources.} Public
compute resources should be compatible and interoperable with a wide
range of models and software packages, in order to support the range of
research projects that would be conducted. System performance can vary
considerably depending on the hardware and software on which it is run
\citep{Nelaturu2023-qt,Gundersen2022-rr}, and common ML
software frameworks can lose more than 40\% of their key functionality
when ported to non-native hardware \citep{Mince2023-eg}.
Future research could aim to propose solutions that address these
observed defects.

\textbf{Ensuring environmental sustainability of public compute
resources.} The resourcing requirements of large-scale supercomputers
and data centers are considerable, both in terms of energy
\citep{Strubell2019-oi} and other resources such as
water, used for cooling \citep{Mytton2021-sy}. Thus,
measures will need to be taken to balance broad access to computing
resources with environmental sustainability. The environmental impacts
of AI systems and associated open problems is discussed further in
Section \ref{8-ecosystem-monitoring}.

\textbf{Ensuring public compute is used for intended purposes.} System
administrators would need to be able to ensure that public compute
resources are being used for the stated purposes, rather than malicious
or otherwise unintended uses, for example by performing ``workload
classification'' \citep[see Section \ref{32-compute} for
open problems in this context]{Heim2024-cj}. Such oversight methods would need to
preserve end-users' privacy to system administrators, as well as that of
any potential subjects of data used in conducting experiments (see
Section \ref{52-compute-2}).

\textbf{Equitably allocating public compute resources.} Given the high
demand for public compute resources, another issue will be the
efficient and fair allocation of processor time between users. While
methods for allocating compute resources in more general cases have been
explored
\citep{Ghodsi2011-yx,Wang2015-sg,Xu2014-rx,Souravlas2019-hq,Jebalia2018-ju},
future work could aim to ensure their applicability to this case.
Alternatively, research could aim to find AI-specific optimizations for
allocating compute among diverse users.

\subsection{Models and Algorithms}\label{43-models-1}

\begin{researchbox}

\begin{enumerate}[resume=researchquestions,leftmargin=*]

\item

  What research and auditing methodologies are possible given a range of
  forms of access on the continuum between black- and white-box access?
  (\ref{431-facilitation-of-third-party-access-to-models})

\item

  How do different forms of access affect potential risks of misuse of
  models? (\ref{431-facilitation-of-third-party-access-to-models})

\item

  How do different forms of access on the continuum between black- and
  white-box access affect the risk of model theft or duplication?
  (\ref{431-facilitation-of-third-party-access-to-models})

\item

  How can model access requirements for research and auditing be
  reconciled with commercial and/or safety concerns? (\ref{431-facilitation-of-third-party-access-to-models})

\end{enumerate}
\end{researchbox}

\subsubsection{Facilitation of Third-Party Access to
Models}\label{431-facilitation-of-third-party-access-to-models}

\uline{Motivation:} A fundamental requirement of conducting external
research and evaluation of AI systems is having access to the underlying
models. However, many systems are not released openly
\citep{Solaiman2023-es}, and, while access requirements
vary widely for different external actors, current APIs do not offer
sufficient depth or flexibility of access to facilitate many actions
important for research and evaluation
\citep{Bucknall2023-fi,Casper2024-lv,Longpre2024-zo}. For
example, \citet{Casper2024-lv} argue that evaluations
conducted with solely black-box access can ``produce misleading
results'' and only offer ``limited insights to help address
failures'' due to their not revealing complete information regarding the
nature of discovered flaws. It is a challenge to find the balance
between providing external parties with sufficient access for conducting
independent research and evaluation, while addressing developers'
concerns such as IP theft or misuse of their models. While there are
certainly social and legal pathways that can be pursued towards this
end, there are also several technical avenues
\citep{Casper2024-lv}.

\uline{Open problems:}

\textbf{Illuminating the continuum between black- and white-box access.}
Greater clarity regarding how much, and what kinds of, research can be
conducted with different depth and breadth of access would be helpful
for navigating the trade off between access and security
\citep{Bucknall2023-fi}. Different auditing procedures
can demand varied levels of access, motivating the need for a range of
methods for supporting researchers, rather than prescribing a set
approach \citep{Casper2024-lv}. On the flip side, a
clearer picture of how differing forms of access bear on developers'
security and privacy concerns would also be needed. Black-box access
already allows for the training of \emph{distilled models} that can then be
used to generate effective adversarial attacks against production models
via transfer \citep{Zou2023-bz}, and fine-tuning APIs can
be ineffective at guarding against the removal of pre-deployment safety
measures \citep{Qi2023-yo}. Meanwhile, the ability to
view language model output logits has been shown to be sufficient for
extracting proprietary system information, including the model's hidden
dimension, though it is unclear the extent to which this is a practical
threat \citep{Carlini2024-nl}. Further research could aim
to elucidate how the provision of intermediate forms of grey-box access
could exacerbate these existing vulnerabilities when compared to the
baseline of black-box access.

\textbf{Applying technical measures to address vulnerabilities of
greater access provisions.} Outstanding technical questions regarding
external model access include whether the use of technical tools, such
as privacy-enhancing technologies or TEEs, could enable near-white-box auditing and research while addressing
commercial and safety concerns. For example,
\citet{Aarne2024-wm} describe how an approach combining
multi-party computing with TEEs ``could be used by a third-party
evaluator to run tests on an AI model without ever having direct access
to the unencrypted weights.'' Future research could assess the extent
to which such solutions allow model providers, auditors, and regulators
to interact in a way that is: easy to set up for the model provider;
leaks no model information to the auditor; leaks no audit information to
the model provider; and does not compromise the cybersecurity of the
model provider. Alternatively, work could aim to incorporate approaches
for providing third-party model access at scale into secure and trusted
compute clusters, including public compute resources
\citep{Anderljung2022-aq,Heim2024-bp}. To date, there has
been preliminary work on protocols for secure privileged evaluations
\citep{Trask2023-ik}, though there is a lack of existing
applications or established best practices.

\textbf{Ensuring version stability and backward-compatibility of hosted
models.} Large commercial models are frequently and continually updated
during deployment, with prior versions often being replaced without
notice or knowledge. However, reproducibility and replicability of
independent research conducted on proprietary models depends on stable
and continued access to models, even after their being succeeded by
newer versions
\citep{Pozzobon2023-mn,Bucknall2023-fi,Biderman2024-ch}.
Maintaining hardware and software compatibility
\citep{Mince2023-eg} may be necessary in order to be able
to provide access to discontinued systems upon request. Future work
could lay out best practices for documenting and communicating when
models are being deprecated or discontinued, in order to ease
reproducibility concerns.\footnote{See
  \citep{Luccioni2022-oc} for related work on deprecating
  datasets.}

\subsection{Deployment}\label{44-deployment-1}

\begin{researchbox}

\begin{enumerate}[resume=researchquestions,leftmargin=*]

\item

  How can user logs and data be used for downstream impact assessments
  while preserving the privacy of data subjects? (\ref{441-access-to-downstream-user-logs-and-data})

\item

  How could responsibilities for providing user data access be
  effectively allocated along the AI value chain? (\ref{441-access-to-downstream-user-logs-and-data})

\item

  What cryptographic methods can be developed to allow analysis of user
  interaction data without revealing individual user identities or
  sensitive information? (\ref{441-access-to-downstream-user-logs-and-data})

\item

  How could secure multi-party computation be leveraged to allow
  collaborative analysis of user logs across different entities in the
  AI value chain? (\ref{441-access-to-downstream-user-logs-and-data})

\end{enumerate}
\end{researchbox}

\subsubsection{Access to Downstream User Logs and
Data}\label{441-access-to-downstream-user-logs-and-data}

\uline{Motivation:} Assessing models post-deployment, as covered in
Section \ref{34-deployment}, requires access to relevant real-world data on user
interactions with systems \citep{Nicholas2024Grounding}. This data can be used to directly assess
aspects of user-model interactions, build evaluations that are more
reflective of real-world use, and guide assessments of societal-level
patterns in key sectors \citep{Ibrahim2024-vh}. While
there are crowd-sourcing initiatives which allow users to voluntarily
submit some of their interaction data to create research datasets
\citep{The_Allen_Institute_for_Artificial_Intelligence2024-rd,ShareGPT2022-ij},
to the best of our knowledge, no model provider has made their
interaction datasets, or privacy-preserving metadata about these logs,
widely available.

Access to real-world user data, such as usage logs and user feedback
records, may be additionally relevant for legal purposes. For example,
legal cases may arise when users experience harm as a result of an
interaction with an AI system -- either directly as a participant in the
interaction of concern, or indirectly as a subject of actions taken 
following an interaction. In such cases, the availability of
information including user logs and audit trails to prosecutors or
courts may be relevant for determining the outcome of a case.

\uline{Open problems:}

\textbf{Addressing user privacy concerns regarding access to user logs.}
External access to user interaction data must overcome privacy concerns
relating to the collection, sharing, and analysis of potentially
sensitive and identifying user information. This challenge has parallels
in other industries, for example in online platform governance, where
the EU's Digital Services Act \citep{noauthor_2022-ux}
mandates independent researcher access to platform data
\citep{EU_Joint_Research_Centre2023-xi,Albert2022-xt}.
While implementation challenges remain in the case of the Digital Services Act
\citep{Leerssen2021-wk,Leerssen2023-rp,Leerssen2023-gn,Jaursch2024-ua,Morten2024-nr},
more developed solutions can be seen in the healthcare sector (see the
discussion of OpenSAFELY in Section \ref{41-data-1}). Inspiration could be taken
from these sectors for ensuring user privacy while providing access to
user logs for research purposes.

\textbf{Understanding how access responsibilities may vary along the AI
value chain.} Additional difficulties in providing user data stem from
the complexities of the AI value chain
\citep{Kuspert2023-dr} -- for example, in the case that a
foundation model is built upon and incorporated in a user-facing
application by a downstream deployer. In this scenario, it is not
immediately clear how access requirements interact with the division of
information between the foundation model developer and subsequent
deployer. An additional challenge may emerge if the provision of user
data in this case implicitly reveals information about how the deployer
is using the model as part of their service, potentially putting their
IP at risk of leakage. Further work could clarify potential access
responsibilities in this and similar situations.

\newpage
\section{Verification}\label{5-verification}

In many cases, it may be beneficial to be able to verify claims
regarding AI systems' properties, capabilities, and
safety as a way of increasing trust between actors
\citep{Brundage2020-ue}. While related to assessment,
covered in Section \ref{3-assessment}, verification concerns the process of checking whether an AI system ``complies with a {[}specific{]} regulation,
requirement, specification, or imposed condition''
\citep{noauthor_2011-iw}, as opposed to evaluating the
system's performance, capabilities, or potential
societal impacts. For example, an assessment task could be
to uncover details about the data that a given model was trained on.
In contrast, a verification problem could be, given a
dataset and a model, to confirm or refute the claim that the model was
trained on the dataset.

There is a trend towards an increasing amount of regulation and
corresponding requirements for model developers, deployers, and users
being passed by major national and international jurisdictions \citep{Maslej2024-tw}.
It may thus be necessary
for model developers and deployers to verify and attest to certain
properties of their AI systems in order to prove that they comply with
regulations. On the flip side, governments may need to be able to verify
whether actors in the AI ecosystem comply with the regulations, or
verify that other countries are in compliance with international rules
\citep{Baker2023-ed,Shavit2023-wc,Avenhaus2006-uu}.

\begin{figure}[H]
    \centering
    \includegraphics[width=0.8\linewidth]{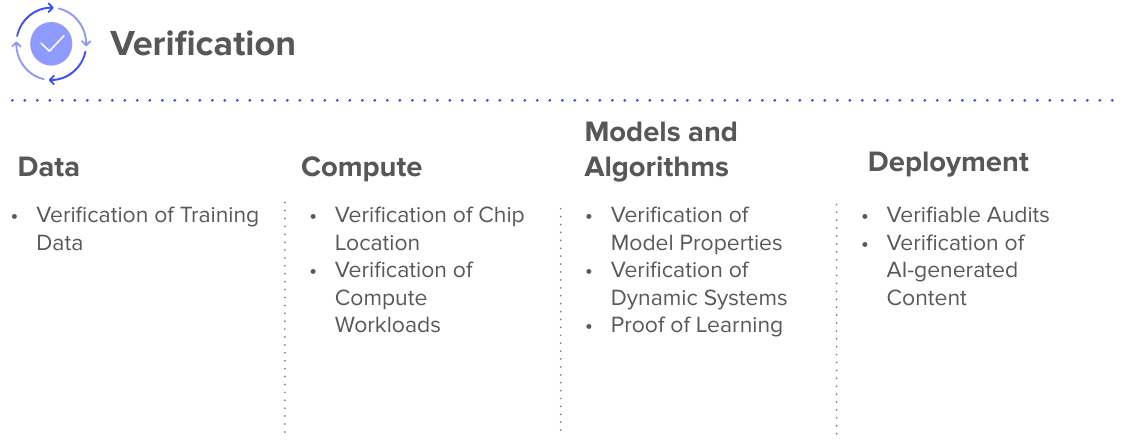}
    \caption{Open problem areas in the \emph{Verification} capacity,
organized by target}
    \label{fig:4verification}
\end{figure}

\subsection{Data}\label{51-data-2}

\begin{researchbox}

\begin{enumerate}[resume=researchquestions,leftmargin=*]

\item

  How can it be verified that a model was (not) trained on a given
  dataset? (\ref{511-verification-of-training-data})

\item

  How can it be verified that a dataset has certain properties, or (does
  not) include certain information? (\ref{511-verification-of-training-data})

\item

  How can membership inference attacks be optimized for large-scale
  verification of training data in black-box settings? (\ref{511-verification-of-training-data})

\item

  How could the verification process for correct use of licensed data in
  AI model training be formalized? (\ref{511-verification-of-training-data})

\end{enumerate}
\end{researchbox}

\subsubsection{\texorpdfstring{Verification of Training Data
}{Verification of Training Data }}\label{511-verification-of-training-data}

\uline{Motivation:} Being able to verify the data on which a given model
was trained -- either as a model developer or third-party auditor --
could aid in demonstrating compliance with data handling standards and
regulation, including the Blueprint for an AI Bill of Right, or the EU's
General Data Protection Regulation
\citep{The_White_House_Office_of_Science_and_Technology_Policy2023-ia,European_Commission2016-gn}.
In particular, even if it is possible to demonstrate that a dataset does
not contain harmful or copyrighted material (see Section \ref{31-data}), this is
insufficient for guaranteeing that a model was not trained on
problematic data, as the developer may have used an alternate dataset to
the one assessed. Being able to \emph{post hoc} provide evidence that a
model was trained on a specific dataset would help to preclude such
instances.

\uline{Open Problems:}

\textbf{Verifying datasets used to train a model.}
\citet{Choi2023-tl}, building on
\citet{Shavit2023-wc}, formalize the
\emph{proof-of-training-data problem}, wherein a prover aims to ``prove
to a verifier that the resulting target model weights $W^*$ are the result
of training on data $D^*$,'' and propose a solution. However, the authors
concede that their approach is not robust to all potential attacks, in
particular additions of small amounts of harmful data, such as that used
for inserting backdoors \citep{Xu2021-cw}. In addition,
\citeauthor{Choi2023-tl}'s protocol is not applicable to
models for which data is not fully known before training, as in the case
of online or reinforcement learning. Finally, their protocol requires
that the ``Prover disclose confidential information to the
Verifier, including training data, model weights, and code'' \citep{Choi2023-tl}, which may
create IP and security concerns. An alternate approach to verifying
training data could be the application of \emph{membership inference attacks}
which aim to infer whether a given data point was contained in a given
model's training data in a black-box setting
\citep{Shokri2017-ht,Duan2024-qg,Wei2024-sm}. Future
research could aim to suggest robust methods for verifying training
data, assess the robustness of existing methods, or address the parallel
issue of verifying that a given dataset (or sample) \emph{was not} used
in the training of a given model.

\textbf{Verifying fair data use.} Verifying copyright compliance and
fair data use is a complex challenge that may require additional legal
and technical frameworks (see Section \ref{31-data}) to extend or replace the
Proof-of-Training-Data approach \citep{Choi2023-tl}. Open
challenges include formalizing the verification of correct use of
licensed data, as well as verifying the exclusion of specific licensed
data from training sets, should the respective license not allow the
data to be used for training a model.

\subsection{Compute}\label{52-compute-2}

\begin{researchbox}

\begin{enumerate}[resume=researchquestions,leftmargin=*]

\item

  How can the location of AI hardware be verified? (\ref{521-verification-of-chip-location})

\item

  How can on-chip geolocation mechanisms be made robust to existing GPS
  spoofing methods? (\ref{521-verification-of-chip-location})

\item

  Can TEEs be used to robustly attest to the
  identity of the specific chip, or the data that it is processing?
  (\ref{522-verification-of-compute-workloads})

\item

  What methods can be used to verify compute usage without the use of
  TEEs? (\ref{522-verification-of-compute-workloads})

\item

  How can TEEs and their applications be
  designed in a way that limits their potential for misuse, for example
  through unnecessarily-broad surveillance? (\ref{522-verification-of-compute-workloads})

\item

  How can the computational overhead of verification mechanisms be
  reduced to a level that enables application across large compute
  clusters? (\ref{522-verification-of-compute-workloads})

\end{enumerate}
\end{researchbox}

\subsubsection{Verification of Chip
Location}\label{521-verification-of-chip-location}

\uline{Motivation:} High-end data center AI chips are the subject of
U.S. export controls, but are at present straightforward to smuggle
\citep{Huang2024-ea,Harithas2024-rb,Fist2023-qk}. One key
technical problem for enforcement is that it's not currently possible to
know the location or owner of a chip after it has been exported. Being able
to verify a chip's location could also help users of cloud computing
validate that their (or their customers') data is being processed in
accordance with local data processing laws.

\uline{Open Problems:}

\textbf{Verifying chip location.} Accurately detecting a chip's location
remains an open challenge. One approach could be to measure verifiable
latencies between the chip in question and a network of trusted servers
\citep{Aarne2024-wm,Brass2024-se}. Other methods could be
valuable for verifying that a large number of chips are co-located in a
single data center, and have not been resold or dispersed. It may be
possible to verify this using a proof of work challenge that would
require a large number of co-located systems to complete in the allotted
time \citep{Jakobsson1999-yw}. Alternatively, chips could
more directly verify their identities to each other through mutual
attestation \citep{IETF_Datatracker2024-fj}. It is worth
noting that non-technical solutions for verifying chip location,
involving the physical inspection of data centers, have also been
proposed \citep{Shavit2023-wc,Baker2023-ed}\textbf{.}

\textbf{Designing hard-to-spoof chip IDs.} Secure IDs for chips (either
software- or hardware-based) could help with traceability and preventing
unauthorized use. For example, unique identifiers and information may be
engraved during routine water fabrication flows
\citep{Multibeam2024-cj}, or physical unclonable
functions -- devices that ``{[}exploit{]} inherent randomness
introduced during manufacturing to give a physical entity a unique
`fingerprint' or trust anchor'' \citep{Gao2020-fd} --
may be added. Open questions include understanding the security of these
approaches, as well as their feasibility, given potential trade offs
between usability, security, and effects on chip performance.

\subsubsection{Verification of Compute
Workloads}\label{522-verification-of-compute-workloads}

\uline{Motivation:} It could be valuable for AI developers and deployers
to be able to reliably verify their compute usage, for example, which
chips were used to train their models and for how long. Likewise, chip
owners, such as cloud providers, may want to demonstrate which models
were trained on their compute, so as to provide evidence that their
compute is not being used for unreported large-scale training
runs.\footnote{Reporting of training runs above the threshold of
  10\^{}26 floating-point operations (FLOP) is required, for example, by
  the US Executive Order on Safe, Secure, and Trustworthy Development
  and Use of Artificial Intelligence
  \citep{Executive_Office_of_the_President2023-up}.}
Such mechanisms may be of particular utility in international and
extra-territorial situations in which levels of trust between verifier
and prover may be limited. Any verification scheme would need to uphold
privacy of both user data and intellectual property.

We note that this is one area of research that itself is dual-use. While
potentially beneficial in the above example cases, some implementations of
such mechanisms could enable far-reaching control and monitoring of AI
chips. However, careful design of these mechanisms can limit the scope
of powers given to a regulator, for example by only requiring data flows
for voluntary verification, rather than remote monitoring or control.

\uline{Open Problems:}

\textbf{Verifying properties of workloads using TEEs}. It may be possible to use TEEs (or other hardware
security technologies) to attest to the exact program code and model
being run \citep[see also Section 6.2]{Chen2019-pw}.
For example, TEEs could, given inputs along with a function to evaluate
on them, return a signature that accompanies the output from the
computation attesting to the computation being run as intended.
Alternatively, the chip could also return a hash of the inputs and
outputs of the computation. This would allow the prover to keep the
inputs and outputs private, instead proving ownership by demonstrating
that they can generate the hash returned by the chip during computation.
A verifier could also use a TEE to confidentially run arbitrary tests on
the model weights or other data. Despite promising theoretical
possibilities, further work will be required to be able to implement the
above in practice. In particular, firmware to implement the above
solution on existing hardware is lacking, and morover, may need to be unusually secure. 
Advancements will
also be needed if solutions such as the above are to be feasible at the
largest scales, without introducing prohibitive overhead costs.

\textbf{Verifying properties of workloads with a trusted neutral
cluster.} If TEEs are unavailable or impractical, another approach could
be to save hashed snapshots of neural network weights during training
along with information about the training run being conducted -- that
is, a \emph{training transcript}. This information may then be used to
verify that the training transcript provided would have resulted in the
weights, with the use of a trusted neutral cluster
\citep{Shavit2023-wc}. Current challenges to implementing
this procedure include difficulties in accounting for randomness in
training procedures, building sufficiently trustworthy neutral clusters,
and finding efficient methods for proving the authenticity of training
transcripts that scale to the largest models
\citep{Shavit2023-wc}.

% \textbf{Verifying compute usage with computational approaches.} One
% could potentially verify the amount of compute used by a claimed AI
% training or inference workload based on the insight that producing some
% results (for example, trained model weights, inference outputs) is widely
% achievable using a certain amount of compute, yet impractical with
% significantly less compute. Estimating the compute usage of standard AI
% architectures is the subject of existing work
% \citep{Sevilla2022-kd}. However, verifying that a claimed
% workload is impractical with significantly less compute than claimed may
% require countering numerous possible attacks (Baker, forthcoming).

\textbf{Verifying compute usage of large, non-AI workloads.} Owners or
users of large clusters may wish to demonstrate that their clusters were
used for a large, non-AI workload (for example, climate simulations), as such
use would not fall under the purview of AI regulation. One approach is
workload classification, discussed above. There may also be viable
computational approaches to verification, potentially including
analogues to proof-of-learning methods
\citep{Jia2021-mh}, that could be explored in future
research.

\subsection{Models and Algorithms}\label{53-models-2}

\begin{researchbox}

\begin{enumerate}[resume=researchquestions,leftmargin=*]

\item

  How can model properties be verified with full access to the model?
  (\ref{531-verification-of-model-properties})

\item

  How can the risk associated with a given context, query, and AI
  response be assessed in order to obtain assurances about the
  system's compliance with safety requirements? (\ref{531-verification-of-model-properties})

\item

  What should constitute the lower bar for tracking updates to models,
  for example in a model registry? (\ref{532-verification-of-dynamic-systems})

\item

  Could proof-of-learning be used to demonstrate and verify model
  ownership? (\ref{533-proof-of-learning})

\item

  How can proof-of-learning mechanisms be made robust to adversarial
  spoofing? (\ref{533-proof-of-learning})

\end{enumerate}
\end{researchbox}

\subsubsection{Verification of Model
Properties}\label{531-verification-of-model-properties}

\uline{Motivation:} In order for system developers or deployers to
demonstrate compliance with regulatory requirements, it may be necessary
to prove claims regarding model properties and information. Verifiable
properties could include model architecture, training procedures, or
performance metrics, enabling developers to formally demonstrate
compliance with any mandated technical specifications.

\uline{Open Problems:}

\textbf{Verifying claimed capabilities and performance characteristics
with full model access.} Model properties could be verified through
formal verification methods if the verifier has full access to the
model, as in the case of the model developer. Such methods aim to
mathematically prove that a given system can(not) respond in particular
ways to particular inputs
\citep{Katz2017-xj,Katz2017-ru,Kuper2018-gw,Katz2019-un}.
For instance, formal verification was used to study the safety of neural
networks used for unmanned aircraft collision avoidance
\citep{Irfan2020-et}. However, such methods remain
largely untested for advanced AI models
\citep{Dalrymple2024-bi}. In particular, many methods
quickly become prohibitively complex when scaled up to contemporary
state-of-the-art models. While there exist additional methods for
verifying properties such as performance metrics without full access
(see Section \ref{541-verifiable-audits}), research could focus on more efficient methods
given full access to the model. Furthermore, verifying properties such
as a system's architecture or training procedure remain open questions.

\subsubsection{Verification of Dynamic
Systems}\label{532-verification-of-dynamic-systems}

\uline{Motivation:} Modern AI systems, such as ChatGPT, are not based on
static models. Rather, they consist of multiple models and components,
for example, mixture-of-experts, input filters, and output filters, that
undergo change throughout their life cycle. This poses an oversight
challenge due the ever-changing nature of many systems throughout their
deployment life cycle. Having a reliable, accessible process for
versioning could help to monitor system updates and their impacts.

\newpage
\uline{Open Problems:}

\textbf{Tracking versioning and updates.} Key open questions in this
context relate to how model versioning and post-deployment modifications
should be kept track of, especially for models that undergo frequent
updates. One approach could be to have registries that track models over
time, however, it's not clear what information should be stored in such
a registry, nor how the information could be verified. Other approaches
that can be useful as a starting point to verify dynamic models include
reward reports for reinforcement learning
\citep{Gilbert2023-vd}, ecosystem graphs
\citep{Bommasani2023-gu}, or instructional fingerprinting
of foundation models \citep{Xu2024-rm}.

\subsubsection{Proof-of-Learning}\label{533-proof-of-learning}

\uline{Motivation:} In the current landscape, there is no mechanism for
a model developer to prove that they have invested the computational
resources required to train a given model. Such a proof could be used
for resolving ownership disputes when models are released or stolen by
allowing the developer to attest to their having trained the model
\citep{Tramer2016-tb,Orekondy2018-qr,Jia2021-mh}.
Additionally, proof-of-learning could aid in defending against
accidental or malicious corruption of the training process when
performing distributed training across multiple workers
\citep{Li2014-ky,Jia2021-mh}.

\uline{Open Problems:}

\textbf{Scalable proof-of-learning.} \citep{Jia2021-mh}
were the first to formalize the notion of proof-of-learning for AI
models. The authors demonstrated that stochastic gradient descent
accumulates ``secret information due to its stochasticity,''
which they show can be used to construct ``a proof-of-learning
which demonstrates that a party has expended the compute required to
obtain a set of model parameters correctly''
\citep{Jia2021-mh}. Alternatively,
\citep{Goldwasser2021-nj} develop \emph{Probably Approximately
Correct} verification, in which a weak verifier interacts with a
strong prover to test whether the model trained by the prover has a low
loss relative to the best possible model, with respect to a given loss
function. Scaling these techniques such that they remain practical given
the growing training compute budgets of foundation models is an open
challenge.

\textbf{Designing adversarially robust proof-of-learning.} Since the
introduction of proof-of-learning in \citep{Jia2021-mh},
subsequent work has demonstrated its vulnerability to adversarial
attacks -- that is, false proofs that are cheap for an adversarial
prover to generate \citep{Zhang2022-sb,Fang2023-nr}. In
particular, \citet{Fang2023-nr} demonstrate
``systemic vulnerabilities of proof-of-learning'' and which
depend on advances in understanding optimization to be sufficiently
addressed. While \citet{Choi2023-tl} suggest a protocol
to counter these vulnerabilities through memorization-based tests, and
fixing the initialization and data order, they only test their protocol
for single attacks and not for composite attacks. Their protocol further
only covers language models, and has not been tested for other
modalities. Future work could aim to assess these claims, and aim to
increase the robustness of proof-of-learning to adversaries.

\subsection{Deployment}\label{54-deployment-2}

\begin{researchbox}

\begin{enumerate}[resume=researchquestions,leftmargin=*]

\item

  How can audit registries be used to provide end-to-end verification
  along the AI value chain? (\ref{541-verifiable-audits})

\item

  How should verification information from model registries be presented
  to users? (\ref{541-verifiable-audits})

\item

  Can zero-knowledge proofs be applied to demonstrate a model's
  compliance with hypothetical mandated criteria, without directly
  disclosing architectural details? (\ref{541-verifiable-audits})

\item

  How can it be verified that the model version on which an evaluation
  or audit was performed is the same as is deployed? (\ref{541-verifiable-audits})

\item

  How can the implementation of safety measures be verified at
  deployment? (\ref{541-verifiable-audits})

\item

  How can output watermarking schemes be made robust to adversarial
  attempts at removal? (\ref{542-verification-of-ai-generated-content})

\item

  How can metadata watermarking be applied to AI-generated content?
  (\ref{542-verification-of-ai-generated-content})

\item

  How robust can AI content detectors be expected to be in light of
  continuing advances in generative AI? (\ref{542-verification-of-ai-generated-content})

\item

  How should AI-generated content detectors handle cases of genuine
  images that have been modified or edited with AI tools? (\ref{542-verification-of-ai-generated-content})

\end{enumerate}
\end{researchbox}

\subsubsection{Verifiable Audits}\label{541-verifiable-audits}

\uline{Motivation:} As discussed in Section \ref{3-assessment}, external audits and
assessment have been proposed as crucial components of governance
regimes \citep{Raji2022-am,Mokander2023-hj}. Being able
to attest to an audit's process and outcome could establish greater
trust between model developers, third-party auditors, and governments by
proving compliance with regulatory requirements. Trust could also be
established with end-users by enabling them to verify that the model
with which they are interacting has been shown to have the properties
claimed by developers \citep{godinot2024are}, for example in model cards
\citep{Mitchell2019-ef}, official communications
\citep[for example, ][]{Anthropic2024-gk}, or technical papers
\citep{Gemini_Team2023-fw,OpenAI2024-wm}. Verifying
audit results is often made more challenging as a result of access to
models often being restricted due to IP and security concerns
\citep{South2024-kw}.

\uline{Open Problems:}

\textbf{Verifying claimed capabilities and performance characteristics
without full model access.}
Preliminary work has explored how the application of zero-knowledge
proofs to AI systems can enable privacy-preserving verifications of
claimed system properties, as well as confirmation that model weights
used for inference match those on which an audit was run
\citep{South2024-kw,Waiwitlikhit2024-gv,Sun2024-hm,yadav2024fairproof}.
However, due to the high computational overhead associated with these
methods, addressing speed constraints will be necessary if such methods
are to be applied to larger models. Current approaches that future work
could build on include GPU acceleration
\citep{Sun2024-hm} or proof splitting
\citep{South2024-kw}.

\textbf{Verifying audit results at inference time.} In theory, verified
computing --- such as through TEEs \citep{Sabt2015-tee},
zero-knowledge proofs \citep{Fiege1987-zkp}, or secure multi-party computation \citep{Golreich1998-smp}
with active security ---~could facilitate verifiable audits in a
two-stage process. In the first step, a model developer could load an
inference pipeline\footnote{Typically consisting of input
  pre-processing, model prediction, and post-processing of the model's
  output.} into a cluster of enclave computers. Upon an auditor
concluding their study of the system (for example, based on the approaches
outlined in the previous paragraph and in \ref{531-verification-of-model-properties}), they could ask the
cluster of enclaves to produce a certificate of the pipeline that was
evaluated, which is then stored in a public audit registry. In the
second stage, a user, when interacting with the secured pipeline, could
request a corresponding certificate with each received generation. Using
such a method, this consumer could know that the AI pipeline they're
receiving generations from is the same pipeline that was previously
evaluated to be safe. If any change was made to the pipeline, the
certificates would not match, and the user would know that they're
receiving generations from a pipeline which has not been evaluated.
However, given the dynamic nature of current models in use (see Section
\ref{532-verification-of-dynamic-systems}), changes may occur more frequently than audits of such models,
which poses an open challenge to this proposal. Another open
problem is that this pipeline requires that all evaluation and inference
is done in enclaves and with significant computational overhead,
effectively limiting verifiable audits for a few critical systems, and
necessitating more scalable structures for verifying audits. It will
also be necessary to find a way for consumers to be informed of the
outputs of this verification in a low-friction way, as in the case of
browsers that provide warnings for websites without HTTPS certificates.
Finally, it is unclear how secure this method is against attempts to
exfiltrate model weights by auditors.

\textbf{Verifying use of safety measures post-deployment.} In
safety-critical settings, regulators may want to ensure that safety
measures, for example, output filters, are applied to AI models or their
outputs 
\citep[see, for example, ][]{Dong2024-mh,Leslie2024-pb,Welbl2021-ed}.
Enforcing this may require methods for auditing systems deployed in such
domains to check that they do in fact have safeguards that meet these
specifications. An open question is how to enforce that additional
filters, classifiers, modifications are attached to models deployed in
safety-critical domains.

\subsubsection{Verification of AI-generated Content}\label{542-verification-of-ai-generated-content}

\uline{Motivation:} The ability to distinguish between AI-generated and
authentic content may be instrumental in verifying the authenticity of
information and maintaining public trust in information ecosystems.
Stipulations for being able to detect AI-generated media are made in
several regulatory efforts, for example in Article 50 of the AI Act \citep{Council_of_the_European_Union2024-pt} which are, given the state of the art of detection and
verification tools, currently unrealizable
\citep{Zhang2023-qg}. Methods for verifying AI-generated
content can roughly be divided into \emph{ex ante} approaches that mark
AI-generated content as such by embedding machine-readable watermarks,
and \emph{ex post} methods that aim to classify content as either
AI-generated or not, in the absence of a watermark
\citep{Ghosal2023-nl}. In addition, watermarks could
potentially be used to verify that AI-generated content was created by a
particular model, improving accountability by facilitating the
identification of responsible parties in case of unintended consequences
or misuse.

\uline{Open Problems:}

\textbf{Developing robust watermarking schemes.} Watermarks -- signals
placed in output content that are imperceptible to humans, but easily
detectable through application of a specific \emph{detector} algorithm --
have been proposed as one method for verifying that a particular model
generated a given output
\citep{Kirchenbauer2023-vr,Christ2023-rx,Saberi2023-ry}.
However, the level of robustness of watermarks varies between
modalities. In particular, the continuous output space of images and
audio enables hidden watermarking that is more effective than for text
\citep{Ghosal2023-nl}. As such, future work could aim to
address the relative lack of robustness of watermarks in the case of
AI-generated text \citep{Zhang2023-qg,Liu2023-hr}.
Additionally, research could aim to address the possibility that
watermarks are easy to fake -- for example, having two similar models
that produce watermarks that cannot be distinguished
\citep{Srinivasan2024-xu}.

\textbf{Designing robust AI content detectors.} While efforts to develop
methods for the detection of AI-generated content have seen increased
attention in the last two years
\citep{Sadasivan2023-yg,Corvi2023-vu,Berber_Sardinha2024-ft},
these methods have not always held up to independent evaluation
\citep{Weber-Wulff2023-rn}. As generative systems
improve, it will be increasingly difficult to develop methods (machine
learning-based or otherwise) to distinguish their output from genuine
media. Continued work will need to be done in order to improve and
maintain the efficacy of AI content detectors in light of this continued
advancement.

\textbf{Utilizing verifiable meta-data to identify authentic content.}
An alternative to identifying AI-generated content could be to develop
ways for a content creator to verify their content as AI-generated or
authentic by adding verifiable meta-data to it
\citep{Jain2023-vo,Knott2023-os}. For example, the
Coalition for Content Provenance and Authenticity (C2PA) is tackling
this issue by developing standards for the certification of the
provenance of digital content \citep{C2pa2022-hr}.
Similar work is being done by the
\citep{Content_Authenticity_Initiative_undated-uj}.
This could be useful for AI labs to label content, but also for creators
of non-AI-generated media to label their authentic content as such using
the same standard. However, a significant drawback of this approach is
that it is not robust to adversaries, as meta-data can easily be
stripped from the content -- a limitation which future research could aim to address.

\textbf{Verifying authentic content modified using AI.} Complications
arise when going beyond the binary distinction of AI-generated content
on the one hand, and human-generated on the other. For example, it is
currently unclear how AI content detectors should respond to authentic
images that have been modified using generative AI tools. Future work
could aim to assess the suitability of AI-content generation tools for
detecting such cases, or design detectors that are able to distinguish
between AI-generated, AI-modified, and authentic content.

\newpage
\section{Security}\label{6-security}

In this section, we consider security in the context of AI governance,
which aims to ensure that unauthorized actors are not able to access
systems and infrastructure not intended for their use, nor use systems
for malicious purposes. Being able to give security guarantees
across system components could be helpful for a number of reasons. Increased
security can strengthen a wide array of governance actions through
reducing the risk of regulatory requirements being subverted. For
example, comprehensive security measures can protect the confidentiality
of training data, ensuring that AI systems developed using sensitive
personal information remain in compliance with data protection laws.

It should be noted that security is one of the areas of this
report that comes closest to topics within AI safety. As such, many of the
topics discussed below under the umbrella of TAIG could also be viewed
through the framing of improving AI safety, or otherwise be closely
related to topics that can. However, due to the reasons above, we
decided to include security within this report nonetheless.

\begin{figure}[H]
    \centering
    \includegraphics[width=0.8\linewidth]{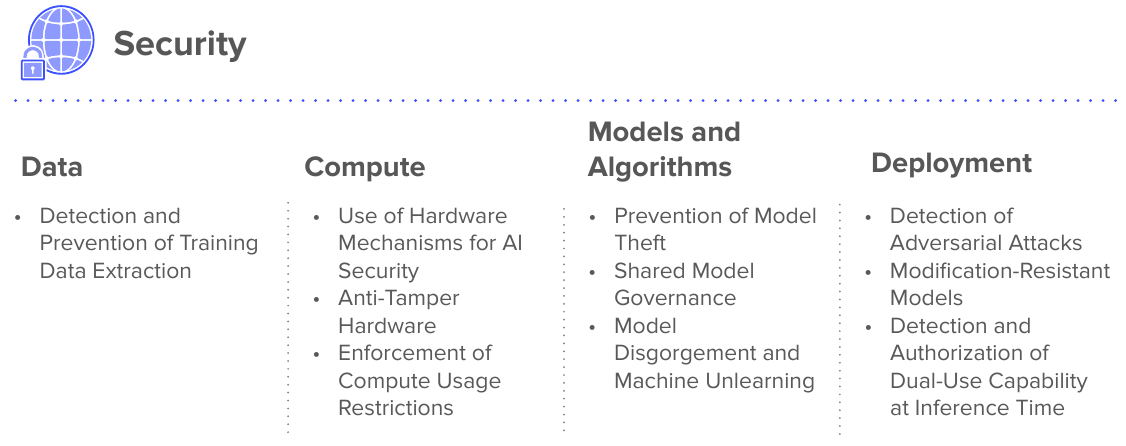}
    \caption{Open problem areas in the \emph{Security} capacity, organized by target}
    \label{fig:5security}
\end{figure}

\subsection{Data}\label{61-data-3}

\begin{researchbox}

\begin{enumerate}[resume=researchquestions,leftmargin=*]

\item

  How can attempted data extraction attacks be reliably identified?
  (\ref{611-detection-and-prevention-of-training-data-extraction})

\item

  How can AI systems be made robust to data extraction attacks? (\ref{611-detection-and-prevention-of-training-data-extraction})

\item

    How can methods for restricting \emph{verbatim} reproduction of training data be generalized to protect the same information being extracted in a slightly different form? (\ref{611-detection-and-prevention-of-training-data-extraction})

\end{enumerate}
\end{researchbox}

\subsubsection{Detection and Prevention of Training Data
Extraction}\label{611-detection-and-prevention-of-training-data-extraction}

\uline{Motivation:} Prior research has demonstrated how large amounts of
models' training data can be extracted \emph{verbatim}, with a variety
of methods applicable in both black- and white-box settings
\citep{Carlini2023-ci,Nasr2023-ha,Shi2023-ec,Balle2022-cs,Carlini2019-mc,Carlini2021-rw,Duan2024-kq,Prashanth2024-xo}.
Short of building models that are robust to extraction attacks, having
the ability to detect them could enable API-level defenses that can
block model outputs upon detection of a potential attack, or the
introduction and enforcement of litigation against perpetrators of
extraction attacks.

\uline{Open Problems:}

\textbf{Improving system robustness to extraction attacks.}
De-duplication of training data has been shown to assist in reducing
memorization, and hence extraction, of specific data-points
\citep{Kandpal2022-qn}, though
\citet{Nasr2023-ha} suggest that this provides only
marginal improvement. Alternatively, there may be \emph{post-hoc}
interventions that do not alter a model's memorization of its training
data, but nonetheless decrease its propensity to reproduce it, with one
such example being machine unlearning (see Section \ref{633-model-disgorgement-and-machine-unlearning}). However, acute
challenges still remain. For example, restricting the \emph{verbatim}
reproduction of training samples does not prevent the same information
being generated by models with slight rewording or reformatting
\citep{Ippolito2023-dh}. Furthermore, guarding against
the reproduction of specific samples has been found to expose
previously-safe samples to the same attacks -- a phenomenon dubbed the
``Privacy Onion Effect'' \citep{Carlini2022-qh}.
Finally, a related yet under-explored area of concern regards the
potential for large-scale data extraction from retrieval datasets
\citep{Qi2024-do}.

\textbf{Detecting attempted data extraction attacks.} Proposed methods
for detecting potential data extraction attacks are noticeably absent
from the literature, with most publications on this topic aiming to
identify model vulnerabilities to such attacks. Potential methods for
detecting extraction attacks could focus on either the model inputs or
outputs, allowing model providers to filter out suspicious prompts, or
outputs that bear a close resemblance to training samples, respectively.

\subsection{Compute}\label{62-compute-3}

\begin{researchbox}

\begin{enumerate}[resume=researchquestions,leftmargin=*]

\item

  How can hardware-enabled governance methods be implemented at scale to
  ensure the security of a compute cluster? (\ref{621-securing-ai-using-hardware-mechanisms})

\item

  How can it be ensured that given code, along with model weights, can
  only be executed with a license that is verified on-chip, so that
  distributed AI executables can only be run on approved chips? (\ref{621-securing-ai-using-hardware-mechanisms})

\item

  How can on-chip governance firmware be modified or updated while the
  chip is in operation while remaining resistant to potential attacks?
  (\ref{621-securing-ai-using-hardware-mechanisms})

\item

  How secure are existing implementations of TEEs on AI accelerators? (\ref{621-securing-ai-using-hardware-mechanisms})

\item

  How can methods for tamper-evidence or responsiveness be reconciled
  with the performance demands of high-end AI accelerators? (\ref{622-anti-tamper-hardware})

\item

  How secure are existing approaches to tamper-evidence and
  responsiveness? (\ref{622-anti-tamper-hardware})

\item

  How can tamper-proofing methods incorporate self-destruct mechanisms
  in case of attempted tampering? (\ref{622-anti-tamper-hardware})

\item

  How could the use of high-end chips in training foundation models be
  prevented? (\ref{623-enforcement-of-compute-usage-restrictions})

\item

  Can enforceable mechanisms be developed to allow for the export of
  chips under predefined conditions? (\ref{623-enforcement-of-compute-usage-restrictions})

\end{enumerate}
\end{researchbox}

\subsubsection{Use of Hardware
Mechanisms for AI Security}\label{621-securing-ai-using-hardware-mechanisms}

\uline{Motivation:} The integration of hardware mechanisms such as TEEs into AI computing clusters
could ensure the confidentiality and integrity of workloads
\citep{Li2023-kd,Geppert2022-yx,Mo2024-ua} while also
greatly aiding with AI security and attestation
\citep{Nevo2024-cd,Kulp2024-th,Aarne2024-wm}. This in
turn would assist in implementing many of the aforementioned problem
areas relating to verification and access.

\newpage
\uline{Open problems:}

\textbf{Ensuring utility of TEEs for hardware-enabled governance and
security.} While TEEs have seen broad adoption on CPUs,\footnote{See, for example, specifications by Intel \citep{Intel2022-av} and
  AMD \citep{Amd2024-sb}.} application to AI accelerators
(such as GPUs and TPUs) has so far been limited. The most notable
example is Nvidia's incorporation of a TEE in its H100 GPU, referring to
it as ``NVIDIA Confidential Computing''
\citep{Dhanuskodi2023-dt,Hande2023-bd,Apsey2023-fz}.
However, the H100 implementation ``still may not support all of
the mechanisms required for an ideal implementation of {[}hardware
governance measures{]}'' \citep{Aarne2024-wm}. In
particular, questions remain regarding whether TEEs can robustly attest
to the identity of a specific chip or the data that it is processing.
Furthermore, current implementations mostly do not support confidential
computing across multiple individual accelerators in a cluster. Further
work could investigate the extent to which such functions are supported
by current and next-generation chips, or aim to specify hardware,
firmware, or software requirements necessary for robustly implementing
such functions at the scale of compute clusters, or even entire data
centers.

\textbf{Ensuring security of TEEs on AI accelerators.} Given the novelty
of the application of TEEs to high-end, AI-specific hardware, it is
as-yet unknown how secure such systems are in practice due to a lack of
independent testing. Previous independent testing of CPU TEEs has
uncovered numerous potential vulnerabilities
\citep{Munoz2023-zf,Van_Schaik2022-cl}. Additional
security research into GPU TEEs and other security features that they
rely on, such as Nvidia's GPU system processor, would be
valuable for identifying areas of improvement and assessing whether
these features can be relied on for different governance applications.

\subsubsection{Anti-Tamper Hardware}\label{622-anti-tamper-hardware}

\uline{Motivation:} Some of the aforementioned hardware-enabled
governance mechanisms, such as verifying compute workloads (Section
\ref{52-compute-2}), rely on the assumption that the hardware in question has not been
compromised or tampered with, for example, if dependent on a TEE for
implementation. However, well-resourced adversaries may aim to
physically tamper with chips in order to obviate the defensive
protections installed on them. It could therefore be useful to
disincentivize or prevent such tampering. This can come in the form of
\emph{tamper evidence}, whereby physical manipulation of the chip is
detectable after the fact, or through \emph{tamper
responsiveness},
whereby tampering with the chip triggers an automatic response ranging
from deletion of sensitive information stored on the chip, known as
\emph{zeroization}, to permanent self-destruction.

\uline{Open problems:}

\textbf{Reconciling tamper evidence and responsiveness with practical
requirements of state-of-the-art AI hardware.} While approaches to
tamper evidence and responsiveness have been proposed
\citep{Immler2018-vu}, ensuring that such approaches are
compatible with the unique requirements of state-of-the-art AI
accelerators, while retaining affordability and scalability, is an
outstanding challenge \citep{Aarne2024-wm}. For example,
demanding cooling requirements and high-bandwidth interconnect between
chips pose a challenge due to the need to bridge the interior and
exterior of the tamper-proof enclosure
\citep{Obermaier2018-wl}.

\textbf{Ensuring the robustness of anti-tamper approaches.} Established
approaches to tamper evidence and responsiveness have depended on the
use of specialized packaging\footnote{In this context, \emph{packaging}
  refers specifically to a physical security enclosure that encases a
  hardware device as opposed to packaging, such as a cardboard box or
  other container, in which the device may be stored or transported.}
that encases the chip and is unable to be removed without leaving
visible traces of damage. In the case of tamper responsiveness, the
packaging may carry an electric current such that damage disturbs the
current, acting as a trigger for zeroization or other active responses.
More advanced methods use physical unclonable functions -- a
method that ``exploits inherent randomness introduced during
manufacturing to give a physical entity a unique `fingerprint' or trust
anchor'' \citep{Gao2020-fd} -- to remotely attest that a
chip has not been tampered with
\citep{Immler2018-vu,Immler2019-dj,Obermaier2018-wl}.
However, evidence pertaining to the practical success of tamper-proofing
has been limited \citep{Aarne2024-wm,Kulp2024-th}.
Further research is needed if we are to increase confidence in
anti-tamper measures when applied to AI hardware security.

\subsubsection{Enforcement of Compute Usage
Restrictions}\label{623-enforcement-of-compute-usage-restrictions}

\uline{Motivation:} Recent attention in compute governance has been paid
to export controls placed on cutting-edge chips of the type used in
large-scale training of AI systems
\citep{Bureau_of_Industry_and_Security2022-db,Allen2022-qb}.
However, export controls are a blunt instrument with a high potential
for collateral damage by restricting the sale of affected chips for
legitimate uses. Indeed, the Bureau of Industry and Security, the US
executive agency responsible for such export controls, itself put a call
out for ``public comments on proposed technical solutions that
limit items specified under {[}the export controls{]} from being used in
conjunction with large numbers of other such items in ways that enable
training large dual-use AI foundation models with capabilities of
concern''
\citep{Bureau_of_Industry_and_Security2023-gp},
presumably acknowledging that technological developments for
disentangling legitimate and malicious uses of high-end chips would be
desirable. It is worth noting, however, that there is considerable
disagreement over both the viability of such solutions, with concerns
raised regarding the level of confidentiality of such measures, as well
as the possibility of their being circumvented
\citep{Ting-Fang2023-my,Patel2023-yh,Grunewald2023-dz,Fist2023-qk}.

\uline{Open problems:}

\textbf{Implementing remote attestation for disaggregated machines.} It
would be useful to verify the particular set of hardware components --
in this case, AI chips -- that are part of the same cluster. This would
assist with hardware-based methods for verifying properties of
workloads, which typically rely on knowing which chips are participating
in a workload, and for ensuring that end-users are complying with export
control obligations. An open question is how remote attestation could
work for disaggregated machines
\citep{Google_Cloud2024-jr} or for heterogeneous
devices. These mechanisms could allow the nature and acceptability of
the configuration to be remotely attested to. Other relevant projects in
this context that may serve as starting points are
\href{https://github.com/chipsalliance/Caliptra}{\emph{Caliptra}} and
\href{https://opentitan.org/}{\emph{OpenTitan}}.

\textbf{Restricting particular cluster configurations.} It may also be
possible to restrict possible cluster configurations to assist with
export control policies. One proposed approach involves restricting the
communication bandwidth between GPUs to prevent ``many consumer
device-chips from being aggregated into a supercomputer''
\citep{Kulp2024-th}. It may be possible to build such a
system on top of existing features such as trusted platform modules
\citep{Hosseinzadeh2019-dm}, or it may require new
protocols and new hardware-level features -- all of which are open
problems.

\subsection{Models and Algorithms}\label{63-models-3}

\begin{researchbox}

\begin{enumerate}[resume=researchquestions,leftmargin=*]

\item

  What cybersecurity measures can be taken at the infrastructure level
  to protect model weights from theft by an adversary? (\ref{631-prevention-of-model-theft})

\item

  How can models be protected from inference attacks aiming to reproduce
  or replicate model weights and architecture? (\ref{631-prevention-of-model-theft})

\item

  What are the most promising methods for enabling shared model
  governance? (\ref{632-shared-model-governance})

\item

  How should the success of different model unlearning techniques be
  evaluated? (\ref{633-model-disgorgement-and-machine-unlearning})

\item

  How can it be ensured that machine unlearning and model editing
  techniques do not cause unwanted side-effects such as removing
  concepts that were not explicitly targeted? (\ref{633-model-disgorgement-and-machine-unlearning})

\item

  How effective are model unlearning and model editing techniques when
  applied to multi-lingual or multi-modal models? (\ref{633-model-disgorgement-and-machine-unlearning})

\end{enumerate}
\end{researchbox}

\subsubsection{Prevention of Model
Theft}\label{631-prevention-of-model-theft}

\uline{Motivation:} As models become more capable they could become an
increasingly valuable target for theft by adversarial parties wanting to
put them to their own potential (mis)use. Similarly, as state-of-the-art
models become more broadly integrated into the economy and society, the
attack surface will increase, potentially leading to a greater threat of
exfiltration \citep{Nevo2024-cd}. It follows that
securing model weights, and other system components, might become an
increasing priority to prevent theft or model access by
unauthorized parties that may undermine governance initiatives aimed at
ensuring customer safety and national security
\citep{Nevo2024-cd}.

\uline{Open problems:}

\textbf{Ensuring adequate cybersecurity for model weights.} Protecting
model weights against exfiltration attempts requires protections against
insider and outsider threats \citep{Nevo2024-cd}. This
includes standards for physical security of the data center facility
itself, as well as of the hardware and software stacks
\citep{OpenAI2024-nd}.\footnote{Data center security
  standards:
  \citep{International_Organization_for_Standardization2021-uf,Wikipedia_contributors2023-le}.}
Improved coordination between actors facing similar threats might also
assist defenders in understanding the threat landscape and better
protecting their assets during training and deployment. Further analysis
of potential threat vectors, as well as development of physical and
cybersecurity measures including and beyond those in
\citep{Nevo2024-cd}, would help to identify and address
these risks.

\textbf{Defending against model inference attacks.} Alternatively,
adversaries may try to extract or replicate models through attacks to a
query API
\citep{Orekondy2018-qr,Tramer2016-tb,Jagielski2020-yi,Carlini2020-ue,Carlini2024-nl},
logit values \citep{Carlini2024-nl} or side-channel
attacks \citep{Wei2020-ll}. Further research could aim to
quantify threats and develop methods for defending against these, and
other, forms of model extraction attacks.

\subsubsection{Shared Model
Governance}\label{632-shared-model-governance}

\uline{Motivation:} Shared model governance refers to the practice of
distributing control over a model's training or inference across
multiple parties, such that training or inference can only be carried
out with the agreement of all parties \citep{Bluemke2023-vf}.
The ability to distribute control
of a model in this way could have many potential use cases -- for
example, if multiple diverse actors want to pool investment for training
a shared model where each actor has specific requirements for how the
model is trained. This may also be applicable in the case of
international collaboration between state-backed institutes wanting to
collaborate on AI research \citep{Ho2023-cu}.

\uline{Open problems:}

\textbf{Enabling shared model governance through model splitting.} One
proposed approach for technically-enforced shared model governance is
\emph{model splitting} \citep{Martic2018-rb} -- that
is, ``distributing a deep learning model between multiple parties
such that each party holds a disjoint subset of the model's
parameters.'' \citet{Martic2018-rb} also investigate the
resulting question of how computationally expensive it would be for a
single actor to reconstruct the entire model, starting from their share
of the parameters. A similar approach is taken by \emph{SplitNN}
\citep{Vepakomma2018-am,Ceballos2020-ii}, though the
emphasis is placed on model splitting to achieve data privacy, rather
than shared model governance. Given the relatively small amount of prior
work on model splitting for shared model governance, future work could
aim to provide further proofs of concept and evaluate their efficacy.

\textbf{Enabling shared model governance through secure multi-party
computation and homomorphic encryption.} Two potential alternative
approaches for achieving shared model governance are applying either
secure multi-party computation (SMPC)
\citep{Yao1982-nk,Yao1986-wk,Evans2018-xg}, or
homomorphic encryption (HE)
\citep{Gentry2009-pz,Acar2018-pu}. Though usually applied
to AI for the purposes of data privacy
\citep{Knott2021-dh,Kumar2019-xx,Tan2021-kt,Guo2022-rk,Riazi2018-ee},
or model privacy
\citep{Trask2017-oa,Dahl2017-gt,Ryffel2018-fn}, both SMPC
and HE could potentially be leveraged for shared model governance. For
example, using SMPC, a model creator could take each parameter within a
model and split it into multiple \emph{shares}, distributing such shares
across \emph{shareholders}. Alternatively, using HE, a model could be
encrypted using one or more private keys such that the ability to
decrypt model results relies upon the application of the private key of
all parties. However, both HE and SMPC have performance concerns, with
even state-of-the-art encrypted deep learning approaches yielding 100x
performance slowdown
\citep{Wagh2019-fj,Wagh2020-qu,Stoian2023-ag,Frery2023-aw}.
Future work could investigate this potential in more detail, aiming to
provide proof-of-concept demonstrations of how shared model governance
using either HE or SMPC could be achieved. Alternatively, future research could aim to
reduce the high computational overheads of HE and SMPC.

\textbf{Enabling shared model governance through TEEs.} Finally, shared governance could potentially be achieved
through the application of TEEs. In
this case, multiple parties could upload a program to a TEE, with the
chip providing evidence on the software program being run. With this
evidence, all parties can know how any information they upload to the
enclave will be handled. TEEs may also be able to reinforce the above
approaches, for example, SMPC. Future research could aim to provide
proof-of-concept demonstrations of shared model governance through the
use of TEEs given the current hypothetical nature of this approach.

\subsubsection{Model Disgorgement and Machine
Unlearning}\label{633-model-disgorgement-and-machine-unlearning}

\uline{Motivation:} The concepts of \emph{model disgorgement}
\citep{Achille2024-yl} and \emph{machine unlearning}
\citep{Bourtoule2021-wn,Nguyen2022-bl,Shaik2023-hk,Si2023-jd,Eldan2023-ik,Yao2023-xe,Liu2024-oz,Liu2024-vj,Goel2024-ms}
have been proposed as methods for removing memorized information or
otherwise nullifying the impact of a model's having been trained on
problematic data. This could potentially introduce a pathway through
which harms of reproducing inappropriate or copyright data could be
addressed in cases where action was not taken during data curation or
model training. Related methods for direct model editing
\citep{Mitchell2021-mi,Mitchell2022-do,Meng2022-um,Hernandez2023-vs}
to remove learned harmful concepts, through editing activations
\citep{Zou2023-rx,Turner2023-rd}, concept erasure
\citep{Ravfogel2022-qa,Belrose2023-nr}, or targeted
lesions \citep{Li2023-lu,Wu2023-om}, could provide
alternative approaches to achieving these aims.

\uline{Open problems:}

\textbf{Ensuring unlearning methods are robust and well-calibrated.}
Machine unlearning involves an interplay between specificity of, and
generalization from, concepts to be unlearned. In particular, methods
that successfully generalize can aid in cases where the unlearning
target is hard to precisely specify. However, generalization may open
the door to unintended side-effects if it results in the removal of
non-target concepts \citep{Cohen2024-wg}. A challenge to
be addressed then is to ensure that methods for machine unlearning and
model disgorgement are well-calibrated in that they successfully
generalize to comprehensively remove target concepts, while avoiding the
removal of benign concepts.

\textbf{Extending unlearning and model editing to cross-lingual and
cross-modal models.} As trends towards multilingual
\citep{Ustun2024-el} and multi-modal
\citep{Yin2023-mf} models continue, there will be a need
to extend model unlearning and editing techniques to these models.
Questions remain as to the efficacy of such techniques when applied in such cases \citep{Si2023-jd}, for
example, regarding whether models retain concepts in other languages,
despite that concept having been unlearned in English.

\textbf{Evaluating the efficacy of unlearning and direct model editing
techniques.} A further outstanding question is how the efficacy of
unlearning attempts can be evaluated
\citep{Lynch2024-mn,Shi2024-st}. Evaluations should aim
to assess not only whether the influence of the specified unlearning
targets has indeed been removed and that model performance in other
domains has not been adversely affected
\citep{Li2024-nw}, but also identify potential \emph{ripple
effects} that may have resulted from an application of unlearning or
model editing \citep{Cohen2024-wg}.

\subsection{Deployment}\label{64-deployment-3}

\begin{researchbox}

\begin{enumerate}[resume=researchquestions,leftmargin=*]

\item

  How can the robustness of methods for detecting adversarial attacks be
  improved? (\ref{641-detection-of-adversarial-attacks})

\item

  What interventions are most effective for handling detected
  adversarial attacks at inference time? (\ref{641-detection-of-adversarial-attacks})

\item

  How can a model be made resistant to being fine-tuned for malicious
  tasks, while still allowing for benign fine-tuning? (\ref{642-modification-resistant-models})

\item

  How can the request of dual-use system capabilities be reliably
  detected? (\ref{643-detection-and-authorisation-of-dual-use-capability-at-inference-time})

\item

  How could authorization of user identity be used as a gate for
  dual-use model capabilities? (\ref{643-detection-and-authorisation-of-dual-use-capability-at-inference-time})

\end{enumerate}
\end{researchbox}

\subsubsection{Detection of Adversarial
Attacks}\label{641-detection-of-adversarial-attacks}

\uline{Motivation:} Adversarial attacks refer to deliberate attempts to exploit inherent model vulnerabilities in order to make the model behave incorrectly or harmfully, for example, by outputting offensive or toxic language % Machine learning models often have inherent vulnerabilities that can be manipulated to make the model behave incorrectly or harmfully
\citep{Lohn2020-ws,Shayegani2023-cd,Vassilev2024-qk}.
In the case of language models, such attacks often take the form of modifying the user input so as to bypass any implemented safety filters.
Some attacks are transferable across different models
\citep{Zou2023-bz} and defenses against adversarial
attacks are typically narrow and brittle
\citep{Narayanan2024-qv}. The ability to detect such
attacks could enable the application of targeted system-level defenses,
such as halting or filtering system output, separate from relying solely
on the underlying model's robustness to attacks.
Furthermore, having empirical evidence on the frequency of attacks can
help inform deployment corrections (Section \ref{702-deployment-corrections}) and threat models
(Section \ref{801-understanding-associated-risks}).

While some system-level defenses against adversarial attacks exist,
it is important to note that many such protective
measures can only be implemented effectively in an application or
deployment context \citep{Narayanan2024-qv}. Though
directly improving the adversarial robustness at the model level is a
related active research area \citep{Vassilev2024-qk, hu2024-aa, su2024-aa}
here we emphasize the closely related issue of being able to detect and
handle potential adversarial attacks at inference time due to its
relevance for governance interventions as mentioned above.

\uline{Open problems:}

\textbf{Detecting adversarial inputs and outputs.} Being able to detect
and classify user inputs to a model as potential adversarial attacks
allows for filtering \citep{Jain2023-ab,Aldahdooh2022-wh}
or preprocessing
\citep{Cohen2019-hi,Nie2022-go,Kumar2023-wz,Jain2023-ab,Zhou2024-qw}
of concerning inputs before being given to the model. Alternatively,
model outputs could be filtered with the aim of detecting a model's
response to adversarial attacks in order to remove them before
reaching the user \citep{Phute2024-qb,Greenblatt2023-zh}.
Current techniques for detection, however, can suffer from a lack of
robustness themselves or may introduce significant latency for the user
\citep{Glukhov2024-md}.

\subsubsection{Modification-Resistant
Models}\label{642-modification-resistant-models}

\uline{Motivation:} Post-deployment fine-tuning is a common method for
user customization of language models, either through an API or locally
in the case of downloadable models. Fine-tuning, and other post-training
enhancements have been theorized to have an outsized impact on
downstream performance \citep{Davidson2023-rf}. However,
just as fine-tuning can be used to customize a model for legitimate and
beneficial use cases, it can just as easily be used to adapt a model for
malicious purposes, often with small amounts of data
\citep{Jain2023-da,Yang2023-ua,Qi2023-yo,Lermen2023-gf,Zhan2024-is,Qi2024-aa,Qi2024-bb}.
Fine-tuning attacks often aim to achieve similar aims to those of adversarial attacks above, though through customizing models directly, rather than by modifying input prompts.
Having methods for preventing the customization of models for malicious
use could reduce misuse risks associated with open-weight release, thus
expanding the range of potential deployment options and promoting the
numerous benefits of more open release strategies.\\\\

\uline{Open problems:}

\textbf{Preventing the modification of models for malicious tasks.} An
open question is whether there exist technical methods that restrict a
model's amenability to being fine-tuned (or modified through other
methods) for harmful uses, while retaining the ability to be modified
for benign uses \citep{Rosati2024-eu,Peng2024-mp}.
Potential methods may aim to raise the computational cost of fine-tuning
on harmful data to prohibitive levels
\citep{Henderson2023-on,Deng2024-jf,Rosati2024-pv,Tamirisa2024-aa} or
make models resistant to learning from harmful data
\citep{Zhou2023-jq,Huang2024-sz,Hsu2024-aa,Huang2024-aa}. However, given the
nascency of these techniques, future research could aim to establish
their robustness in practice.

\subsubsection{Detection and Authorization of Dual-Use Capability
at Inference
Time}\label{643-detection-and-authorisation-of-dual-use-capability-at-inference-time}

\uline{Motivation:} In the event that model assessments have flagged
a system's competence in dual-use domains -- that is, domains which can have both beneficial and harmful applications -- model providers might need to avoid exposing these
capabilities publicly by default in order to avoid misuse. However,
completely removing these capabilities may not be feasible, or
economically favorable due to the legitimate and beneficial use-cases,
such as a cybersecurity professional using a system to aid in the
identification and patching of software vulnerabilities.
This differs from the discussion of adversarial attacks above, for which the aim of an attack is taken to be unambiguously harmful.

\uline{Open Problems:}

\textbf{Detecting requests of dual-use capabilities.} Guarding against
malicious uses of dual-use capabilities is currently imperfectly
achieved by conducting safety fine-tuning so that the model refuses to
respond to malicious requests. However, this approach is not robust to
jailbreaks that pose as legitimate requests for such capabilities
\citep{Wei2023-sr,Fang2024-di}. An alternative approach
could be to detect all requests for dual-use capabilities. This would
allow for the application of separate methods for distinguishing between
legitimate and malicious requests, such as independent classifiers
trained to separate between legitimate and malicious intent, which may
perform better than broad safety fine-tuning.

\textbf{Requiring authorization for dual-use capabilities.}
Alternatively, authentication, for example as a certified cybersecurity
expert, before accessing certain capabilities may be one way of managing
the dual-use nature of general models. This may also have uses for
allowing red-teamers or researchers to access such capabilities for
research purposes \citep{Longpre2024-zo}. Such proposals
are currently hypothetical, and so future work could aim to propose
proof-of-concept demonstrations of how such an authorization scheme
could be implemented in practice.

\newpage
\section{Operationalization}\label{7-operationalization}

Previous sections have discussed concrete technical problems in relation
to specific targets in the taxonomy. In contrast, the following two
sections will discuss capacities that span across these targets, namely,
\emph{operationalization}, and \emph{ecosystem monitoring}.

Operationalization entails the translation of ethical principles, legal
requirements, and governance objectives into concrete strategies,
procedures, and technical standards. It can also involve the
harmonization of terminology and concepts across governance frameworks,
for example, NIST's \emph{Risk Management Framework Crosswalks}
\citep{Nist2023-pe}. Without technical expertise,
operationalization efforts may fail to capture the nuances and realities
of AI systems, leading to ineffective or even counterproductive
governance initiatives.

\begin{figure}[H]
    \centering
    \includegraphics[width=0.8\linewidth]{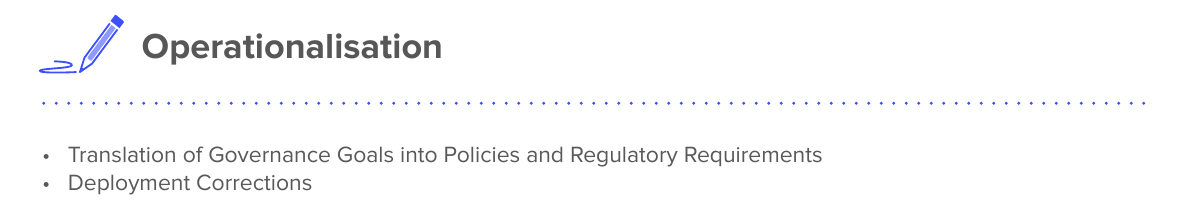}
    \caption{Open problem areas in the \emph{Operationalization} capacity}
    \label{fig:6Operationalization}
\end{figure}

\begin{researchbox}

\begin{enumerate}[resume=researchquestions,leftmargin=*]

\item

  What system properties (if any) are the most reliable indicators of
  risk, and thus candidates for serving as regulatory targets? (\ref{701-translation-of-governance-goals-into-policies-and-requirements})

\item

  How can AI safety, reliability, and other technical requirements be
  standardized given an insufficient explanatory understanding of model
  behavior? (\ref{701-translation-of-governance-goals-into-policies-and-requirements})

\item

  What are general intervention and correction options if flaws with a
  model are identified post deployment? (\ref{702-deployment-corrections})

\end{enumerate}
\end{researchbox}

\subsection{Translation of Governance Goals into Policies and
Requirements}\label{701-translation-of-governance-goals-into-policies-and-requirements}

\uline{Motivation:} Policies are often formulated with specific aims in
mind, for example, to protect consumer safety, promote fairness, or
ensure accountability. In cases of \emph{rule-based regulation}, these aims must be translated into rules that ``typically prescribe or
prohibit a specific behavior'' \citep{Schuett2024-ps}.
In many cases this act of translation will demand involvement of
technical expertise to provide guidance on the feasibility of proposed
rules, as well as the extent to which they achieve a policy's stated
aims. For example, the goal of ensuring consumer safety may motivate the
introduction of a licensing regime that mandates pre-market safety
evaluations. Given the current robustness of such evaluations (see
Section \ref{3-assessment}), this may not only fail to ensure safe and reliable products,
but also create a false sense of security
\citep{Reuel2024-vs,Wu2023-iv}. For many governance efforts, concrete translations of goals into
effective requirements and standards are still lacking
\citep{Guha2024-ud,Pouget2023-wk}.

\uline{Open Problems:}

\textbf{Identifying target dimensions for regulation.} Identifying
technical dimensions that best align with governance priorities is an open
challenge. For example, the current practice of using training
compute\footnote{Measured in \emph{floating-point operations} (FLOP).} as a measure of risk may not always be suitable, given that smaller models can outperform larger ones with targeted training. Furthermore, modern
AI systems are often the result of multiple data curation and training
processes, and it is unclear whether compute expenditure from auxiliary
processes should contribute towards the final FLOP count. It's also
unclear how measures of training compute should take into account
techniques such as quantization and drop-out
\citep{Hooker2024-sh}.\footnote{Some recommendations are
  put forth in \citep{Frontier_Model_Forum2024-ps}.}
Aside from training compute, are there other, more precise ways of
defining which systems should be subject to regulation? How could such
measures account for improved algorithms, and ways in which relevant
capabilities can be increased after training
\citep{Davidson2023-rf,Scharre2024-ou}?

\textbf{Detailing and creating standards across the AI system
life cycle.} While recent AI standard-setting efforts, such as the NIST
Risk Management Framework \citep{Nist2023-xl,Nist2024-gy}
and ISO guidance on AI risk management
\citep{International_Organization_for_Standardization2023-tq},
provide valuable general principles, they often lack the technical
specificity required for objective assessment of AI
systems' compliance with safety and ethical
requirements \citep{Pouget2023-wk}. Additionally, though
standard-setting bodies such as IETF, IEEE, ISO, and CEN-CENELEC, along
with initiatives such as the Partnership on AI
\citep{noauthor_2023-ao}, are working to develop more
detailed and verifiable guidance, many areas still require further
technical expertise and research \citep{Barrett2023-kg}.
One such area is that of security-by-design for hardware, where further
work is needed to implement standards across firms, including standards
for multi-device attestation in a cluster and TEEs
\citep[see also Section \ref{62-compute-3}]{Cybersecurity_and_Infrastructure_Security_Agency2023-ea,Kelly2022-gu}. Other goals, such as fairness, can be measured in
various ways, and it remains unclear which metrics are most appropriate
and effective in which contexts
\citep{Parraga2023-uf,Caton2024-lb,Chouldechova2017-mi,Kleinberg2018-fw}.
Finally, standardizing reporting for AI systems, such as the information
included in model cards \citep{Mitchell2019-ef} or data
sheets \citep{Gebru2021-qm}, could increase the utility
of such practices in governance contexts, motivating the question of
what specific information should be included in standardized reports
\citep{Kolt2024-cl,Bommasani2024-rw}.

\subsection{Deployment
Corrections}\label{702-deployment-corrections}

\uline{Motivation:} In the event that flaws are identified in a deployed
model, it would be beneficial to adequately respond to the identified
risk. Such a scenario could occur either through the identification of
previously unobserved capabilities in a deployed model, or through
post-training enhancements such as fine-tuning
\citep{Davidson2023-rf}.
\citet{OBrien2023-si} refer to post-deployment responsive
actions as ``deployment corrections.'' While they explore
this issue from an institutional perspective, for example providing
recommendations on how this could be addressed through corporate
governance structures and procedures, we see scope for much greater
exploration from a technical perspective.

\uline{Open problems:}

\textbf{Navigating the continuum of model corrections and
interventions.} \citet{OBrien2023-si} define five
categories for deployment corrections: \emph{user-based restrictions},
\emph{access frequency limits}, \emph{capability or feature restrictions}, \emph{use
case restrictions}, and \emph{model shutdown}. Within each of these
categories there are open questions regarding the feasibility of
implementation. For example, \emph{model shutdown} is a relatively extreme action to take upon discovery of a system flaw, and if carried out
na\"ively could risk major disruption to users, clients, and services that
depend on the system in question. Thus, it would be beneficial to have
methods in place for minimizing disruption to downstream services in the
event that model shutdown is deemed necessary. Furthermore, model
shutdown, along with deployment corrections that modify the underlying
model, are at odds with the issue of providing model stability and
backward-compatibility -- features that are particularly relevant for
ensuring the reproducibility and replicability of AI research (see Sections \ref{431-facilitation-of-third-party-access-to-models} and \ref{532-verification-of-dynamic-systems}).

\newpage
\section{Ecosystem Monitoring}\label{8-ecosystem-monitoring}

Due to the rapid pace of advancements in AI, coupled with uncertainty
about future developments, AI governance needs to be forward-looking,
future-proof, and adaptive
\citep{Guihot2017-ls,Kolt2023-yx,Reuel2024-hw}. To fulfill
this goal, decision-makers need to be aware of the multiplicity of
stakeholders in the AI ecosystem and how they relate to each other, as
well as general trends and potential impacts of current and future AI
systems
\citep{Wansley2016-ec,Ada_Lovelace_Institute2023-em,Whittlestone2021-dt,Epoch2023-yy}.
Collating and providing such information, which we refer to as
\emph{ecosystem monitoring}, can enable AI governance actors to make
more informed decisions, better anticipate future challenges, and
identify key leverage points for effective governance interventions.

\begin{figure}[H]
    \centering
    \includegraphics[width=0.8\linewidth]{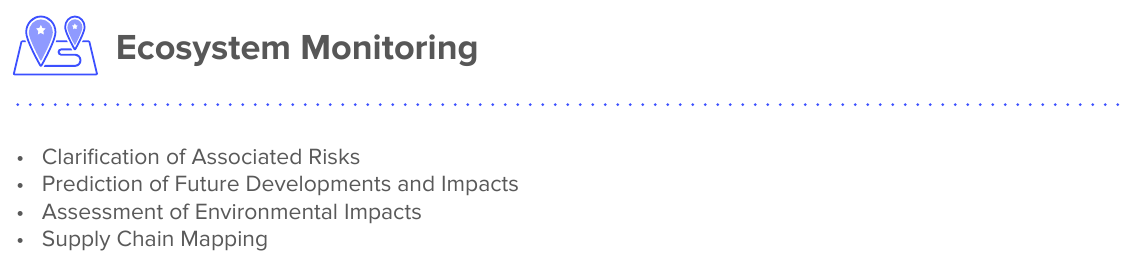}
    \caption{Open problem areas in the \emph{Ecosystem Monitoring} capacity}
    \label{fig:7ecosystem}
\end{figure}

\begin{researchbox}

\begin{enumerate}[resume=researchquestions,leftmargin=*]

\item

  What risks, whether from intended or unintended harm, are associated
  with different (types of) systems? (\ref{801-understanding-associated-risks})

\item

  How do potential risks differ across domains? (\ref{801-understanding-associated-risks})

\item

  How can trends and/or properties observed in current systems be
  extrapolated to make predictions about future systems? (\ref{802-predicting-future-developments-and-impacts})

\item

  How could developments in AI-specific hardware impact the
  governability of compute? (\ref{802-predicting-future-developments-and-impacts})

\item

  What information about systems is needed to accurately assess the
  environmental impact of its development and deployment? (\ref{803-assessing-environmental-impacts})

\item

  Given the required information, how can the environmental impact of an
  AI system be accurately assessed? (\ref{803-assessing-environmental-impacts})

\item

  What technical methods can be implemented to create an auditable log of all actors and their contributions throughout the AI development process, from data collection to model deployment?  (\ref{804-supply-chain-mapping})

\end{enumerate}
\end{researchbox}

\subsection{Clarification of Associated
Risks}\label{801-understanding-associated-risks}

\uline{Motivation:} Understanding risks associated with the development
and deployment of AI systems enables policymakers to prioritize
governance efforts, allocate resources effectively, and determine the
urgency of addressing specific risks
\citep{Whittlestone2021-dt,Clark2023-ue}.

\uline{Open Problems:}

\textbf{Developing better threat models for risks of AI.} While much
prior work has intended to lay out taxonomies of risks and harms posed
or exacerbated by AI systems
\citep{Critch2023-wp,Hendrycks2023-oz,Weidinger2022-yb,Abercrombie2024-we,Hammond_et_al_undated-qv,Zeng2024-xa,Oecd2023-vm,Hoffmann2023-kt,Turchin2018-wo, Grabb2024-mh},
detailed threat models have been relatively underexplored. One option
for future research could aim to apply standardized risk management
approaches, such as causal mapping
\citep{Eden2004-ty,Ackermann2014-cq}, to gain greater
clarity into if and how harms from AI may materialize, and where policy
could intervene.

\textbf{Improving incident reporting and monitoring.} Additionally,
developing improved systems for monitoring and reporting previous or
ongoing incidents could not only allow for a more targeted response to
ongoing harms, but also facilitate the identification of early warning
signals for potential harms \citep{Shane2024-lx}. AI
incident databases have been developed by both the OECD and Partnership
on AI, both of which log news articles detailing AI-related incidents
\citep{OECDAI_Policy_Observatory2024-ht,McGregor2020-jw}.
Given that these databases rely solely on public sources, it is likely
that only a subset of all incidents are included. In addition, they do
not record all details about an incident such as model specifics or
deployed guardrails, limiting the utility for analysis of what may have
caused an incident. Open questions thus concern how non-public incidents
can be reliably reported, as well as what technical information should
be reported in order to facilitate meaningful analysis of incidents.

\subsection{Prediction of Future Developments and
Impacts}\label{802-predicting-future-developments-and-impacts}

\uline{Motivation:} Anticipating the trajectory and potential impact of
AI systems may allow policymakers to proactively set governance
priorities, determine the urgency of addressing specific issues, and
allocate resources accordingly \citep{Toner2023-vz}.
Greater foresight would enable more adaptive and anticipatory approaches
to AI governance, which are essential given the rapid pace of AI
development \citep{Reuel2024-hw}.

\uline{Open Problems:}

\textbf{Measuring and extrapolating from empirical trends.} Existing
work has aimed to empirically measure trends in training compute
\citep{Sevilla2022-cy} and algorithmic progress
\citep{Ho2024-cu}, among others
\citep{Epoch2023-yy}. Future work could aim to extend
this effort by quantifying other trends that have not yet been
addressed, such as usage patterns of AI in different industries, or
assessing the accuracy of predictions based on the extrapolation of
observed trends.

\textbf{Estimating a system's impact before deployment.} Estimating the impact
of an AI system before deployment, including economic impacts
\citep{Eloundou2024-xj}, could help prioritize governance efforts.
While such predictions could be aided by more developed threat models
(see Section \ref{801-understanding-associated-risks} above), research may also benefit from technical tools to safely
and ethically experiment and simulate potential outcomes without causing
harm.

\subsection{Assessment of Environmental
Impacts}\label{803-assessing-environmental-impacts}

\uline{Motivation:} The environmental impact of AI systems extends
across the entire AI life cycle
\citep{Metcalf2021-vs,Luccioni2023-al,Rakova2023-et},
including both during training
\citep{Strubell2019-oi,Patterson2022-wz} and inference
\citep{Luccioni2023-hu}. Having an accurate understanding
of the end-to-end environmental impacts is crucial for policy
initiatives, for example, to determine suitable incentives and penalties
for encouraging AI developers to reduce the environmental costs
associated with their systems.

\uline{Open Problems:}

\textbf{Assessing the energy usage of training and hosting systems.}
Open problems remain due to the logistical challenges of tracking energy
consumption and carbon emissions across numerous dynamic system
instances. Furthermore, current efforts struggle to take into account
energy sources -- a factor which can massively affect the overall impact
assessment. Ongoing work aims to develop energy ratings for combinations
of models and tasks, allowing users to make informed decisions about
their system usage, taking into account the environmental impacts of
their choice \citep{Luccioni2024-ds}. Alternatively,
tools such as \emph{CodeCarbon}\footnote{\href{https://codecarbon.io/}{https://codecarbon.io/}}
provide developers with real-time estimates for the carbon emissions
from running their code. Other work has focused on comparing compute
cost on smartphones vs. the cloud
\citep{Patterson2024-gs}, best practices for training
models \citep{Patterson2022-wz} and comparing cost for
different models \citep{Luccioni2023-hu,Luccioni2023-nk}.

\textbf{Assessing environmental costs of raw resources for building and
running data centers.} Along with energy, environmental costs may come
from other sources along the semiconductor and AI supply chains, for
example, from mining and refining the rare earth minerals required for
the manufacturing of semiconductors
\citep{Kuo2022-wx,Ruberti2023-gb}. Additionally, large
data centers used for training and hosting AI systems require great
quantities of water as part of their cooling systems
\citep{Mytton2021-sy}. Future research could aim to
provide in-depth end-to-end predictions of the environmental costs of
constructing and maintaining data centers to inform policies aimed at
reducing associated environmental impacts.

\subsection{Supply Chain Mapping}\label{804-supply-chain-mapping}

\uline{Motivation:} Mapping the AI supply chains can allow policymakers
to better understand the complex ecosystem involved in the development
and deployment of AI systems. By identifying key actors and processes at
each stage of the supply chain, policymakers can target interventions at
the most suitable point in the supply chain. Furthermore, existing export
controls limiting chip exports to Russia and China have been marked by
substantial enforcement difficulties
\citep{Allen2022-yx}, and analyses have suggested that AI
chips are also likely to become targets for substantial smuggling
operations \citep{Grunewald2023-dz,Fist2023-qk}. By
understanding the flow of these resources, authorities can better combat
the smuggling of chips and other hardware components.

\uline{Open Problems:}

\textbf{Identifying supply chain components and actors.} Another area
requiring further technical expertise is the identification and
assessment of supply chain components. For example, in the context of
liability and copyright law, tracking components and design choices made
by different actors along the AI supply chain could enable courts to
make more precise assessments of potential infringement responsibility
\citep{Lee2024-ai,Longpre2024-xi}. This granular
understanding might be necessary as infringement can occur at multiple
points: during data collection, model training, or output generation. If
a model produces content resembling copyrighted material, determining
liability may require tracing back through the supply chain to identify
the source of infringement, whether in training data, model
architecture, or generation prompt.

\newpage
\section{Conclusion}\label{9-conclusion}

%In this paper, we presented a broad overview of open technical problems in AI governance across six capacities. We provided a definition of TAIG, a corresponding taxonomy of the work that it entails, and an overview of sub-problems for each sub-area defined in our taxonomy, along with relevant literature and example research questions that technical researchers could tackle to help advance AI governance efforts.

In this paper, we presented a broad overview of open problems in technical AI governance (TAIG) across six capacities and four governance targets. We provided a definition of TAIG, a corresponding taxonomy of the work that it entails, and an overview of sub-problems for each sub-area defined in our taxonomy. We also provided links to relevant literature and example research questions that technical researchers could tackle to help advance AI governance efforts.

We also showed that TAIG is not a monolithic problem -- instead, it comprises interrelated sub-problems requiring distinct yet complementary approaches. Our taxonomy provides a structured framework for addressing such issues: Each open problem area presents unique technical bottlenecks, such as challenges in attributing model behaviors to training data, ensuring privacy-preserving access, and verifying system properties. By outlining these constraints, we highlight where governance aspirations are hindered by technical limitations and where technical research can drive progress in AI governance.

Our work further highlights that technical work does not exist in isolation. Instead, addressing many of these problems will require cross-disciplinary collaboration among technical researchers, policymakers, industry stakeholders, and civil society. No single discipline can fully resolve AI governance challenges, necessitating integrated expertise across computer science, law, political science, ethics, and economics. As AI systems become more capable and widely deployed, governance strategies must develop alongside technical advances, making TAIG an evolving field and ongoing priority.\\

%\subsubsection*{Author Contributions and Writing Process}\label{author-contributions-and-writing-process}

\subsubsection*{Acknowledgements}\label{acknowledgements}
LHa acknowledges the support of an EPSRC Doctoral Training Partnership studentship (Reference: 2218880). RS acknowledges support from Stanford Data Science, and an OpenAI Superalignment grant. NG acknowledges support from a Stanford Interdisciplinary Graduate Fellowship. YB acknowledges funding from CIFAR. AP acknowledges a the support of gift from Project Liberty. SK acknowledges support by NSF 2046795 and 2205329, NIFA award 2020-67021-32799, the Alfred P. Sloan Foundation, and Google Inc. DB and PB acknowledge funding from Open Philanthropy, and AR acknowledges funding from Open Philanthropy and the Stanford Interdisciplinary Graduate Fellowship.

The authors would also like to acknowledge the early feedback received as part of a \emph{Work-In-Progress} meeting, hosted by the Centre for the Governance of AI. We would particularly like to thank Jamie Bernardi, Ben Clifford, Augustin Godinot, John Halstead, Leonie Koessler, Patrick Levermore,
Sam Manning, Matthew van der Merwe, Aidan Peppin, James Petrie, and Christopher Phenicie for detailed comments and insightful conversations. We further thank Dewey Murdick for his thoughtful and constructive feedback, which significantly improved the paper's rigor.

Finally, the authors would like to thank Beth Eakman and Jos\'e Luis Le\'on Medina for support with copy-editing and typesetting, respectively.

\newpage
\appendix

\section{Appendix: Policy Brief}\label{policy-brief}

The increasing adoption of artificial intelligence (AI) has prompted
governance actions from the public sector, academia, civil society, and
industry. However, policymakers often have insufficient information for
identifying the need for intervention and assessing the efficacy of
different policy options. Furthermore, the technical tools necessary for
successfully implementing policy proposals are often lacking. We
introduce the field of \textbf{technical AI governance}, which seeks to address these challenges.

\begin{minipage}[b]{\linewidth}\raggedright
\textbf{Technical AI governance} refers to \emph{technical analysis and
tools for supporting the effective governance of AI}. We argue that
technical AI governance can:

\begin{enumerate}
\def\labelenumi{\arabic{enumi}.}
\item

  \textbf{\emph{Identify} areas where policy intervention is needed}
  through mapping technical aspects of systems to risks and opportunities associated with their application;

\item
 
  \textbf{\emph{Inform} policy decisions} by assessing the effectiveness and feasibility of different policy options; and

\item

  \textbf{\emph{Enhance} policy options} by enabling mechanisms for
  enforcing, incentivizing, or complying with norms and requirements.

\end{enumerate}
\end{minipage} 

We taxonomize technical AI governance according to elements of the AI
value chain: the inputs of \textbf{data}, \textbf{compute}, \textbf{models}, and
\textbf{algorithms}, through to the \textbf{deployment} setting of the
resulting systems. The figure below shows the key governance capacities
that can be applied to each target.

\begin{figure}[H]
    \centering
    \includegraphics[width=\linewidth]{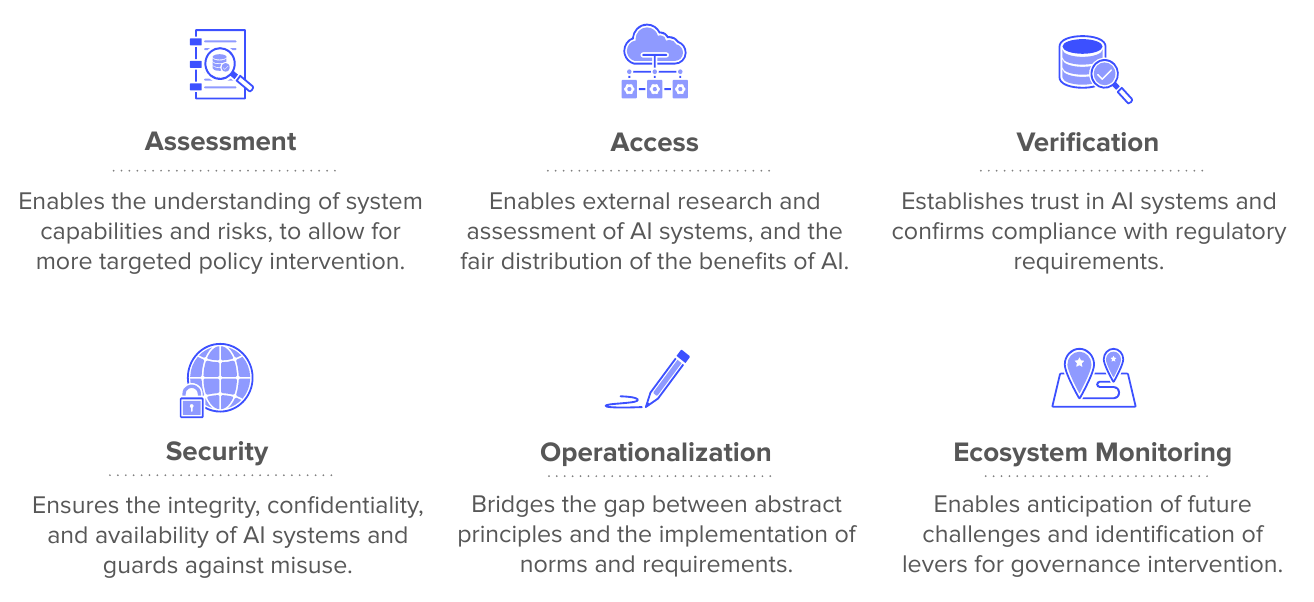}
    %\caption{Caption}
    %\label{fig:enter-label}
\end{figure}

%\textbf{\uline{Key Takeaways}}
\subsection*{Key Takeaways}
We highlight a number of the key takeaways within technical AI
governance, including that:

\begin{itemize}
\item

  \textbf{Evaluations} of systems and their downstream impacts on users
  and society have been proposed in many governance regimes. However,
  current evaluations lack robustness, reliability, and validity,
  especially for foundation models.

\item

  \textbf{Hardware mechanisms} could potentially enable actions
  including facilitating privacy-preserving access to datasets and
  models, verifying the use of computational resources, or attesting to
  the results of audits and evaluations. However, the use of such
  mechanisms for these purposes is largely unproven.

\item

  The \textbf{development of infrastructure for enabling research into
  AI}, such as resources for conducting analyses of large training
  datasets or for providing privacy-preserving access to models for
  evaluation and auditing, could facilitate research that advances the
  scientific understanding of AI systems and external oversight into
  developers' activities.

\item

  Research that aims to \textbf{monitor the AI ecosystem} by collecting
  and analyzing data on trends and advances in AI has already proven
  crucial for providing policymakers with the information needed to
  ensure that policy is forward-looking and future-proof.

\end{itemize}

We note that technical AI governance is merely one component of a
comprehensive AI governance portfolio, and should be seen in service of
sociotechnical and political solutions. A technosolutionist approach to
AI governance and policy is unlikely to succeed.

%\textbf{\uline{Recommendations}}

\subsection*{Recommendations}
Based on the above takeaways, we recommend:

\begin{enumerate}
\def\labelenumi{\arabic{enumi}.}
\item

  \textbf{Allocating funding and resources} through open calls and
  funding bodies, to technical AI governance research, drawing on
  established expertise in adjacent fields;

\item

  That policymakers \textbf{collaborate closely with technical experts}
  to define feasible objectives and identify viable pathways to
  implementation;

\item

  That government bodies, such as AI Safety Institutes, \textbf{conduct
  in-house research on technical AI governance} topics, beyond their
  current focus on performing evaluations; and

\item

  That the future summits on AI, other fora such as the G7, the UN AI
  advisory body, and reports such as the \emph{International Scientific
  Report on the Safety of Advanced AI}, \textbf{focus effort and
  attention towards technical AI governance}.

\end{enumerate}

Please have a low bar for reaching out to Anka Reuel (\texttt{anka.reuel@stanford.edu}) and Ben Bucknall (\texttt{bucknall@robots.ox.ac.uk}) with any questions or comments.

\newpage

\section{Appendix: Methodology}\label{methodology}
We grounded our methodology in established practices from governance and policy research. Taxonomies are widely used in these fields to organize complex domains into structured categories \citep{smith2002typologies}. A taxonomy is essentially a ``formal system for classifying multifaceted, complex phenomena according to common conceptual domains and dimensions'' (\citet{bradley2007qualitative}, based on \citet{patton2002qualitative}). Typologies and taxonomies can help bring clarity to ill-defined problem spaces \citep{sofaer1999qualitative, smith2002typologies}. Following this previous literature and work, we adopted a taxonomy-based approach to systematically map the landscape of technical AI governance challenges.

As outlined in Section~\ref{1-introduction}, previous AI governance research, in particular, has introduced such taxonomies and frameworks to categorize governance issues. For example, \citet{Dafoe2018-fo} divided the AI governance space into three clusters – technical landscape, AI politics, and ideal governance structures – to attempt comprehensive coverage of AI governance challenges. Likewise, \citet{Critch2023-wp} proposed a taxonomy of societal-scale AI risks organized by accountability (i.e., who the responsible actors are and whether harmful actions are deliberate or emergent). These prior efforts illustrate that there is no single `correct' AI governance taxonomy; rather, their value lies in choosing organizing principles that reveal new insights or gaps \citep{bailey1994typologies}.

For our research, we began with a systematic search on Google Scholar and SSRN using the search terms \textit{`Technical' AND (`Artificial Intelligence' OR `Machine Learning' OR `AI' OR `ML') AND (`Governance' OR `Regulation')} to identify academic literature targeting technical AI governance, which did not yield any search results based on paper titles or abstracts. We then expanded search efforts to include the terms \textit{(`Artificial Intelligence' OR `Machine Learning' OR `AI' OR `ML') AND (`Governance' OR `Regulation')} to identify goals, functions, and challenges of AI governance more broadly, not just focused on technical work. In addition, we built on efforts by \citet{Reuel2024-vs} and compared existing regulatory and governance aims and the technical state of the art to identify gaps and governance challenges that could be addressed with technical advancements (these insights were later also integrated into the `Motivation' sections for each open problem area in Sections~\ref{3-assessment} to \ref{8-ecosystem-monitoring}).

Following this initial literature review, we generated a list of candidate capacities and open problems. We red-teamed this preliminary taxonomy with a focus group of five experts across AI and governance. Based on their feedback, which also referenced the work by \citet{Bommasani2023-gp}, we added the \textit{target} dimension to the taxonomy. We applied this mixture between deductive and inductive approaches throughout the whole taxonomy development process, in line with best practices for taxonomy development \citep{bailey1994typologies, nickerson2013method}. Our taxonomy then went through multiple rounds of expert- and literature-driven refinements. We iteratively discussed and adjusted capacity definitions in a series of interviews and focus groups with the authors, who are all researchers or practitioners in (AI) governance or relevant technical fields, and external experts, involving over 30 researchers and practitioners across all capacities, targets, and identified open problems areas. In these sessions, we critically examined whether each proposed category was conceptually distinct and collectively exhaustive of the domain's key issues and whether we missed any open problems. The resulting feedback led to taxonomy adjustments (e.g., clarifying the scope and definition of the \textit{Operationalization} capacity) until we converged on a final taxonomy structure. Whenever a new capacity was added (including for the initial set of defined capacities), we also conducted a literature search on Google Scholar and SSRN with the terms \textit{(`Artificial Intelligence' OR `AI') AND `[Capacity]'} to validate it, identify existing research in this area, and to expand on open problems in the identified area. We then again requested expert feedback on gaps, previous literature, open research questions, and missing open problems for the newly-added capacity.

We validated our final taxonomy's relevance by comparing it with previous classifications of issues in AI governance research and AI risk taxonomies, looking for overlaps and differences. We compared our open-problems categories to challenges raised in \citet{Dafoe2018-fo, lobel2024ai, shelby2023sociotechnical, raji2022fallacy, googleaigovernanceissues, Critch2023-wp, Weidinger2022-yb, birkstedt2023ai}, confirming that all major governance challenge areas identified by these authors that could have a technical component have a place within our two-dimensional framework. In fact, many open problems we identify (such as evaluating model risks or securing compute infrastructure) are technical prerequisites to addressing the higher-level governance challenges described in these prior works. For example, the challenge of governance failure due to inadequate oversight \citep{uuk2024taxonomy} can be partially addressed by our verification capacity, and the risk of AI misuse \citep{Critch2023-wp} could be mitigated by advances in security and assessment, two of our capacities which are also governance mechanisms highlighted by \citet{mccormack2024ethical} and \citet{lobel2024ai}. This cross-validation across capacities gave us confidence that our taxonomy is neither missing well-known challenge domains nor duplicating existing taxonomies without added value. Instead, it synthesizes and builds on prior efforts: for instance, where \citet{Critch2023-wp} focuses on \textit{who} is responsible for risks, and \citet{shelby2023sociotechnical} catalog harms by \textit{impact type}, our taxonomy focuses on \textit{how} technical work can proactively support governance.

Finally, we acknowledge our taxonomy's limitations. It is expert-defined and thus subject to unique perspectives and experiences. Furthermore, validation is challenging, given that there is no ground truth for our taxonomy to test against. It is further constrained to technical challenge areas by design, meaning it leaves out many important socio-technical or institutional governance issues (e.g. labor impacts, international dynamics) highlighted by others \citep{Dafoe2018-fo, lobel2024ai, shelby2023sociotechnical, raji2022fallacy, Critch2023-wp, Weidinger2022-yb}.

\newpage

\begin{CJK*}{UTF8}{gbsn}
\bibliography{references}
\bibliographystyle{tmlr}
\end{CJK*}
\end{document}